\documentclass [epj] {svjour}
\usepackage{graphics}
\usepackage{color}
\usepackage{amsmath,amssymb,bm}
\usepackage{float}
\usepackage{amsmath,amssymb,bm}
\usepackage{graphics}
\usepackage{graphicx}
\usepackage{float}
\newcommand{\betae}{\ensuremath{\beta _{e}}}
\newcommand{\betap}{\ensuremath{\beta _{p}}}

\def\hugeint{ \Huge \mbox{ $ \int $} \normalsize}
\def\hugeslash{ \Huge \mbox{ $ / $} \normalsize}
\newcommand{\dborn}{\ensuremath{t^2  + u^2  - 4 \, M^2 \, ( t+u ) + 6 M^4}}

\def\hugeint{ \Huge \mbox{ $ \int $} \normalsize}

\newcommand{\modvecp}{\ensuremath{|\bm{p} |}}
\newcommand{\modvecq}{\ensuremath{|\bm{q} |}}

\newcommand{\etasurvl}{\ensuremath{\frac{\eta}{v \,\ell }}}
\newcommand{\pietasurvl}{\ensuremath{\frac{ \pi \,\eta}{v \,\ell }}}
\newcommand{\yzero}{\ensuremath{y_0}}
\newcommand{\yun}{\ensuremath{y_1}}
\newcommand{\zun}{\ensuremath{z_1}}
\newcommand{\ydeux}{\ensuremath{y_2}}
\newcommand{\zdeux}{\ensuremath{z_2}}
\newcommand{\ytrois}{\ensuremath{y_3}}
\newcommand{\ztrois}{\ensuremath{z_3}}
\newcommand{\betac}{\ensuremath{\beta _{e}}}

\newcommand{\pmoins}{\ensuremath{p^{-}}}
\newcommand{\pplus}{\ensuremath{p^{+}}}
\newcommand{\qmoins}{\ensuremath{q^{-}}}
\newcommand{\qplus}{\ensuremath{q^{+}}}

\newcommand{\kslash}{\ensuremath{ / \! \! \!     k }}

\newcommand{\pmoinsslash}{\ensuremath{ / \! \! \!    \pmoins }}
\newcommand{\pplusslash}{\ensuremath{ / \! \! \!     \pplus }}
\newcommand{\qmoinsslash}{\ensuremath{ / \! \! \!    \qmoins }}
\newcommand{\qplusslash}{\ensuremath{ / \! \! \!     \qplus }}
\newcommand{\xslash}{\ensuremath{ / \! \! \!     x }}

\newcommand{\vep}{\ensuremath{ {v} _{e^+} }}
\newcommand{\ubarem}{\ensuremath{ \bar{u} _{e^-} }}

\newcommand{\up}{\ensuremath{ {u} _{p} }}

\newcommand{\vbarpbar}{\ensuremath{ \bar{v}_{\bar{p}} }}
\newcommand{\soft}{\ensuremath{\mathrm{Soft}}}

\def\hugeint{ \Huge \mbox{ $ \int $} \normalsize}

\newcommand{\pmoinsslashmM}{\ensuremath{ / \! \! \!    \pmoins - M}}
\newcommand{\pplusslashpM}{\ensuremath{ / \! \! \!     \pplus + M  }}
\newcommand{\qmoinsslashpm}{\ensuremath{ / \! \! \!    \qmoins + m }}
\newcommand{\qplusslashmm}{\ensuremath{ / \! \! \!     \qplus - m }}

\newcommand{\pmoinsslashM}{\ensuremath{ / \! \! \!    \pmoins_{M} }}
\newcommand{\pplusslashM}{\ensuremath{ / \! \! \!     \pplus_{M}  }}
\newcommand{\qmoinsslashm}{\ensuremath{ / \! \! \!    \qmoins_{m} }}
\newcommand{\qplusslashm}{\ensuremath{ / \! \! \!    \qplus_{m}  }}

\begin{document}
\title {Radiative corrections in nucleon time-like form factors measurements }

\author{ Jacques Van de Wiele\inst{1} and Saro Ong\inst{1,2} }

\institute{ Institut de Physique Nucl\'eaire, IN2P3-CNRS, Universit\'e de
Paris-Sud, 91406 Orsay Cedex, France.
\and Universite de Picardie Jules Verne, F-80000 Amiens, France}

\date{Received: date / Revised version: date }

\abstract{The completely general radiative corrections to lowest order, including the final and initial state radiations, are studied in proton-antiproton annihilation into an electron-positron pair.
Numerical estimates have been made in a realistic configuraton of the
 PANDA detector at FAIR for the proton time-like form factors measurements.} 



\titlerunning{Radiative correction in time-like form factors measurements}
\authorrunning{J. Van de Wiele et al.}
\maketitle

\section{Introduction.}

Precise polarization measurements of the proton electromagnetic form factor [1] 
confirm the $Q^2$ dependence up to $Q^2=8.5$ GeV$^2$, of the ratio 
$R=\mu G_E/G_M$ 
(where $\mu$ is the proton's magnetic moment and $G_E$ and $G_M$ are
 the electric and magnetic proton form factors) showing an approximately
  linear decrease of $R$ with  $Q^2$.
This fact is in disagreement with results obtained from a new Rosenbluth
 cross section measurement [2] and suggests that the source of the discrepancy
  is not simply experimental. Recently, there has been a revival of interest 
  in this subject [3]. 
Some theoretical works [4-6] have investigated the two-photon exchange
 corrections to the lowest order QED. This effect has been shown to resolve 
 partially the discrepancy [3,7]. It is well established that the Rosenbluth
  method is much more sensitive to radiative corrections than the 
  polarization method. Until now, intense theoretical activities to evaluate 
  the radiative corrections to elastic electron-proton scattering, 
  which include higher order radiative  corrections [8,9] or 
  model-dependent box diagram calculation [10], incorporating the nucleon's
   substructure, are not able to draw a definitive conclusion on this 
   discrepancy. We should note that two-photon exchange corrections are 
   small in general and at a 1$\%$ level for a large class 
of experiments [3]. 
With this renewal of interest, the importance of theoretical descriptions 
of nucleon form factors in the space-like and also in the time-like region 
is emphasized.\\
In principle, time-like form factors can be evaluated from the space-like
 equivalents by means of dispersion relations.  
  The ratio between electric and magnetic proton form factors
    was recently analysed in the framework of dispersion relations,
   using space-like and time-like data [11,12].  
   However, all the published data in the time-like region [13-16] assumed 
$G_E=G_M$ to hold for all $Q^2$ and 
not only at threshold. Moreover, accurate data at high energy are lacking.
Close to threshold the discrepancy between the LEAR [13] and BaBar [16] data
has to be resolved. Therefore a measurement of the proton form factors in the time-like region is planned at PANDA at FAIR in proton antiproton annihilation into an electron positron pair with unprecedented high accuracy [17].\\

As mentioned above, the radiative corrections may be correctly evalued
 in the kinematical configuration of the scattering experiment to extract
  the physics observables of interest. This paper is devoted to a 
  theoretical investigation of the $\bar{p} p \rightarrow e^+e^-$ process 
  including radiative corrections to lowest order of perturbation theory.
   Among the recent papers devoted to this subject, one should mention [18]
    where the possibility to measure the charge asymmetry is presented
assuming the proton as a point-like particle. The charge-odd part presented 
in the differential cross section is the origin of this asymmetry. 
Recently, the authors of [19]  reevaluated this correction in the laboratory
 frame, omitting the contribution of the hard photon as well as the 
 contribution from the initial state radiation.\\
The first results of full simulations with PANDA detector [17] show 
that precisons of the order of 3$\%$-5$\%$ can be obtained for 
the cross section measurement. 
In this context, we need to evaluate the radiative correction due to
the final state radiation, namely photon emission from the electron
 or the positron, as well as the radiation in the initial state at the 
 proton vertex.

  In this paper, the complete general radiative corrections to lowest order, 
 to the $\bar{p} p \rightarrow  e^+ e^-$ channel are investigated,
 in the kinematical configuration of the planned experiment with
  the PANDA detector  at FAIR. The hard photon contribution from the reaction 
  $\bar{p} p \rightarrow e^+e^-\gamma $ is evaluated with different models
   assumption for the electromagnetic form factors, in contrast to what was 
   done  in [18].
 Indeed, to take into account the radiative corrections in the data analysis,
  we need to know
 the efficiency and acceptance corrections for a complete simulation. 
 This study is a first step towards a correct extraction of the form factors
  from angular distribution measurements
 in the data analysis stage.\\

\section{Electromagnetic nucleon current operator and Born cross section}
Let us first introduce our notations and the definition of the electromagnetic
 nucleon current operator with the magnetic and electric form factors of 
 the nucleon.
 For the annihilation process, in the one photon exchange approximation:
$$\bar{p} (p^-) ~+~p(p^+)  \rightarrow e^+(q^+) ~+~ e^-(q^-)$$
The corresponding Feynman diagram for this reaction is given in Fig.1.
\begin{figure}[H]
\begin{center}
\resizebox{0.35\textwidth}{!}{
\includegraphics{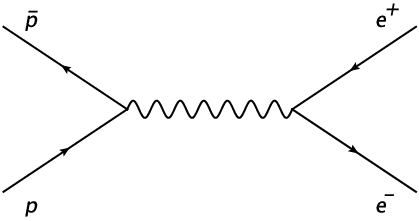}
}
\caption{One photon exchange diagram for the process
 $\bar{p} p \rightarrow  e^+e^-$. }
\label{Fig: 1} 
\end{center}
\end{figure}
The Born amplitude has the form
\begin{equation}\label{eq1}
{\displaystyle M^B ={4 \pi \alpha \over s} \bar{v}(p^-)
\Gamma_{\mu}u(p^+)\bar{u}(q^-)\gamma^{\mu}v(q^+)}
\end{equation}
with 
\begin{equation}\label{eq2}
{\displaystyle \Gamma_{\mu}= F_1(s) \gamma_{\mu}  
 + {F_2(s) \over 4M} [\gamma_{\mu},  \slash  \! \! \! q]  }
\end{equation}
where  M is the proton mass
and the complex quantities $F_1 (s) $ and $F_2(s) $ are the Dirac
 and Pauli form factors respectively. $G_M(s)$ the magnetic and $G_E(s) $
 the electric form factors 
are related to the Dirac and Pauli form factors $F_1$ and $F_2$ by:  
\begin{eqnarray}
G_E(s)
&=&
F_1(s)+\tau F_2(s)    
\label{eq3}  \\[1mm]
G_M(s)
&=&
F_1(s)+ \, F_2(s)
\label{eq4}
\end{eqnarray}
with
\begin{eqnarray*}
&&
s=q^2=(p^-  +  p^+)^2,   \hspace{12mm} \tau=s/(4M^2)
 \\[1mm]
&& \hspace{10mm}
\beta_p^2=1-4M^2/s
\end{eqnarray*}
The differential cross section in the Born approximation in the center of 
mass has the form (neglecting the electron mass $m$):
\begin{eqnarray}\label{eq5}  
\Big[{d\sigma \over d \Omega} \Big]_B
={\alpha^2  \over  4s\beta_p}  \{& & \hspace{-2mm} |G_M(s)|^2(1+\cos^2 \theta)
\nonumber \\[0mm]
\hspace{-1mm}
&  +& \, (1-\beta_p^2)\, |G_E(s)|^2 \sin ^2 \theta \, \}
\end{eqnarray}
\noindent
where $\theta$ is the scattering angle of the positron. This cross section
 was first derived by the  authors of [20]. The analytical form is exactly
 the same for the electron.
 The expression of the differential cross section in terms of the
 electric and magnetic form factors shows that we have access only
 to the modulus of these complex quantities. It is only with a polarized beam
 or a polarized target that we can learn something on their relative phase.
In the particular case, where the proton is considered as a pointlike
 particle, with $G_M(s)=G_E(s)=1$, the formula (\ref{eq5}) reduces to 
\begin{equation}\label{eq6} 
{\displaystyle  \Big [{d\sigma \over d \Omega} \Big]^0_B
={\alpha^2  \over  4s\beta_p}~\left\{ 
2-\beta_p^2  \sin ^2 \theta \right\}} 
\end{equation}
It should be noted that the shape and the normalization of the differential
 cross section is sensitive to the model assumption of the form factors.
 We display in Fig. 2  
the distribution for two different models: the "Babar" model -model 1- which is 
obtained by a fit to the Babar data[16] and the  model 2 which is the model
 of F. Iachello and Q. Wan [21].
\begin{figure}[H]
\begin{center}
\resizebox{0.45\textwidth}{!}{
\includegraphics{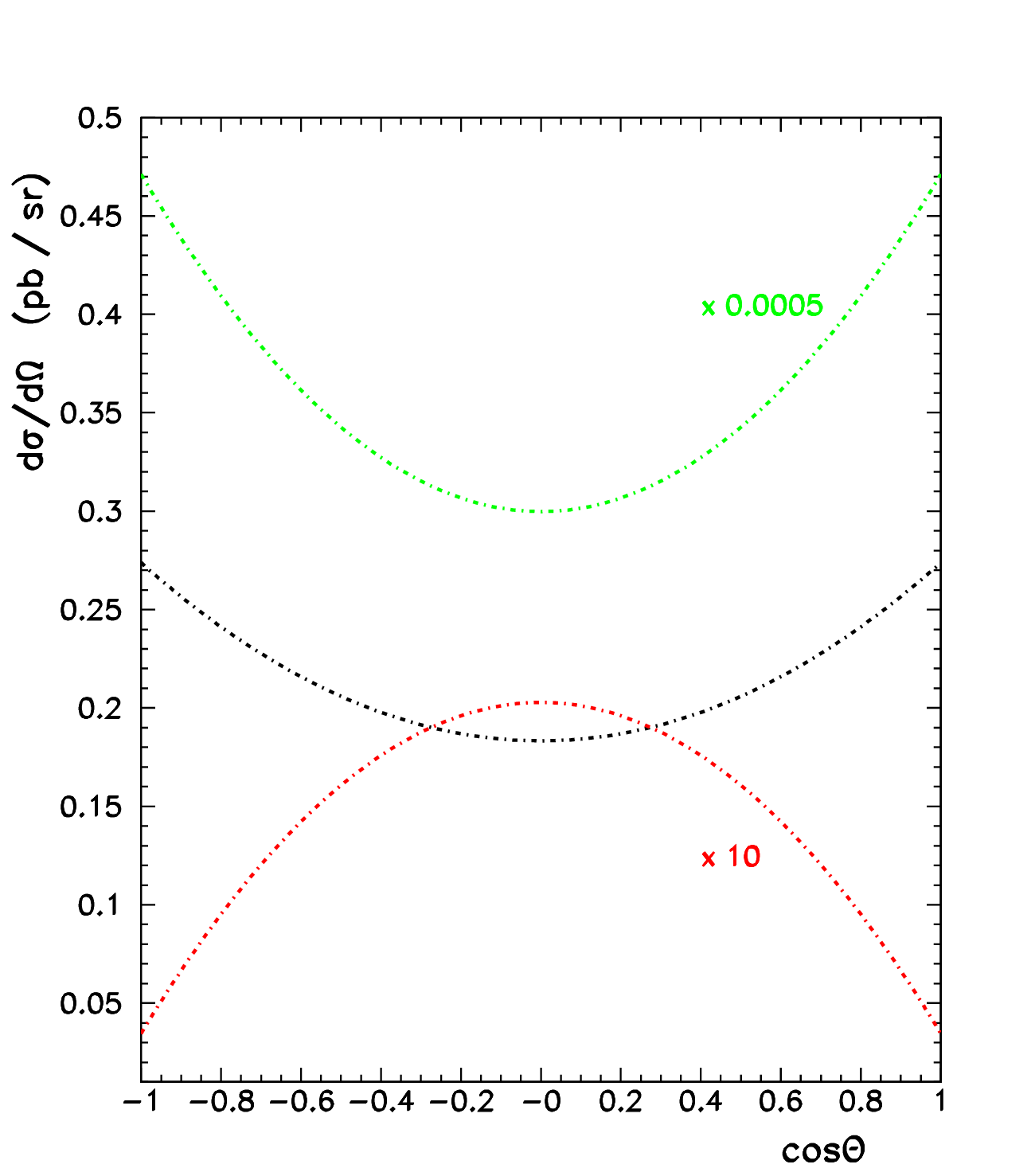}
}
\caption{Born cross section versus $\cos\theta$ for  $ s = 12.9 $ GeV$^2$:
 point-like model of the proton (green dash-dotted line), 
 model 1 (black dash-dotted line) and model 2 (red line)}
\label{Fig: 2} 
\end{center}
\end{figure}  
    
 For this phenomenological model 1,
  the form factors $G_M$ and $G_E$ read:
\begin{eqnarray}\label{eq7}
G_{_M} = \mid G_{_M} \mid e^{i \varphi _{_M}};
\hspace{12mm}
G_{_E} = \mid G_{_E} \mid e^{i \varphi _{_E}}
\end{eqnarray}
According to the pQCD asymptotic behavior in the space-like region,
the modulus of the magnetic form factor has the form:
\begin{eqnarray}\label{eq8}
\mid G_{_M} \mid = 
\frac{A}{ \displaystyle q^4 \,
\big( \ln ^2 \frac{ \scriptstyle q^2}{\scriptstyle \Lambda^2 _{QCD}} 
+ \pi^2 \big)}
\end{eqnarray}
with
\begin{eqnarray}\label{eq9}
\Lambda _{QCD} = 0.3 \,  GeV \hspace{10mm} A =98 \, GeV^4 
\end{eqnarray}
An another important parameter is the ratio $R$ defined as
$R = \mid G_{_E} \mid / \mid G_{_M} \mid  $ which is parametrized by:
\begin{eqnarray}\label{eq10}
R = \frac{\mid G_{_E} \mid }{\mid G_{_M} \mid  }
=
1 + \Big[ \frac{q^2}{4 M^2 } - 0.3\Big]^{-2}    \ln \frac{q^2}{4 M^2 }
\end{eqnarray}
and the relative phase
\begin{eqnarray}\label{eq11}
\varphi _{_E} -\varphi _{_M} =
\pi  \bigg[ \, 1 - e^{ -\frac{1}{2} 
                  \big( \frac{\scriptstyle q^2}
                             {\scriptstyle 4 M^2  } -1 
                  \big) }
            \, \bigg]
\end{eqnarray}
\noindent
The fit to the Babar data are guided by the following constraints:
\begin{eqnarray*}
&1)& q^2 = 4 M^2   \hspace{6mm}    \Rightarrow  R =1                     \\
&2)& q^2 \to \infty  \hspace{9mm}  \Rightarrow   R \to 1                 \\ 
&3)& q^2 = 4 M^2  \hspace{6mm} \Rightarrow    
\varphi _{_E} -\varphi _{_M} = 0    \\
&4)& q^2 \to \infty  \hspace{9mm} \Rightarrow   
 \varphi _{_E} -\varphi _{_M} \to \pi 
\end{eqnarray*}
The constraints $1)$ and $3)$ follow from the definition of the electric and
magnetic form factors in terms of the Dirac $F_1$ and Pauli $F_2$ form factors.
The constraints $2)$ and $4)$ follow from the theorems of Phragm\'en and
Lindel\"{o}f \cite{22} which states that the ratio $ G_{_E}/G_{_M} $
is the same in both the space-like and the time-like regions
when $ q^2 \rightarrow \pm \infty$. In the space-like region, these form
factors are real and they are asymptotically real also in the time-like
region. In the framework of dispersion relations and fitting the available
data both in time-like and space-like regions, the autors of ref[11]
predict the presence of space-like zero of $ G_{_E}/G_{_M} $
at $ q^2=(-11\pm 2 ) $ GeV$^2$.
 The fact that the relative phase tends to $ \pi $ radians
 follows from  the relation as mentioned in formula (12) of ref[11]
 and the presence of the space-like zero ratio.

The shapes of the ratio $R$ versus $q^2$ for the two models
under consideration in this article are displayed in Fig.3.
\begin{figure}[H]
\begin{center}

\resizebox{0.45\textwidth}{!}{
\includegraphics{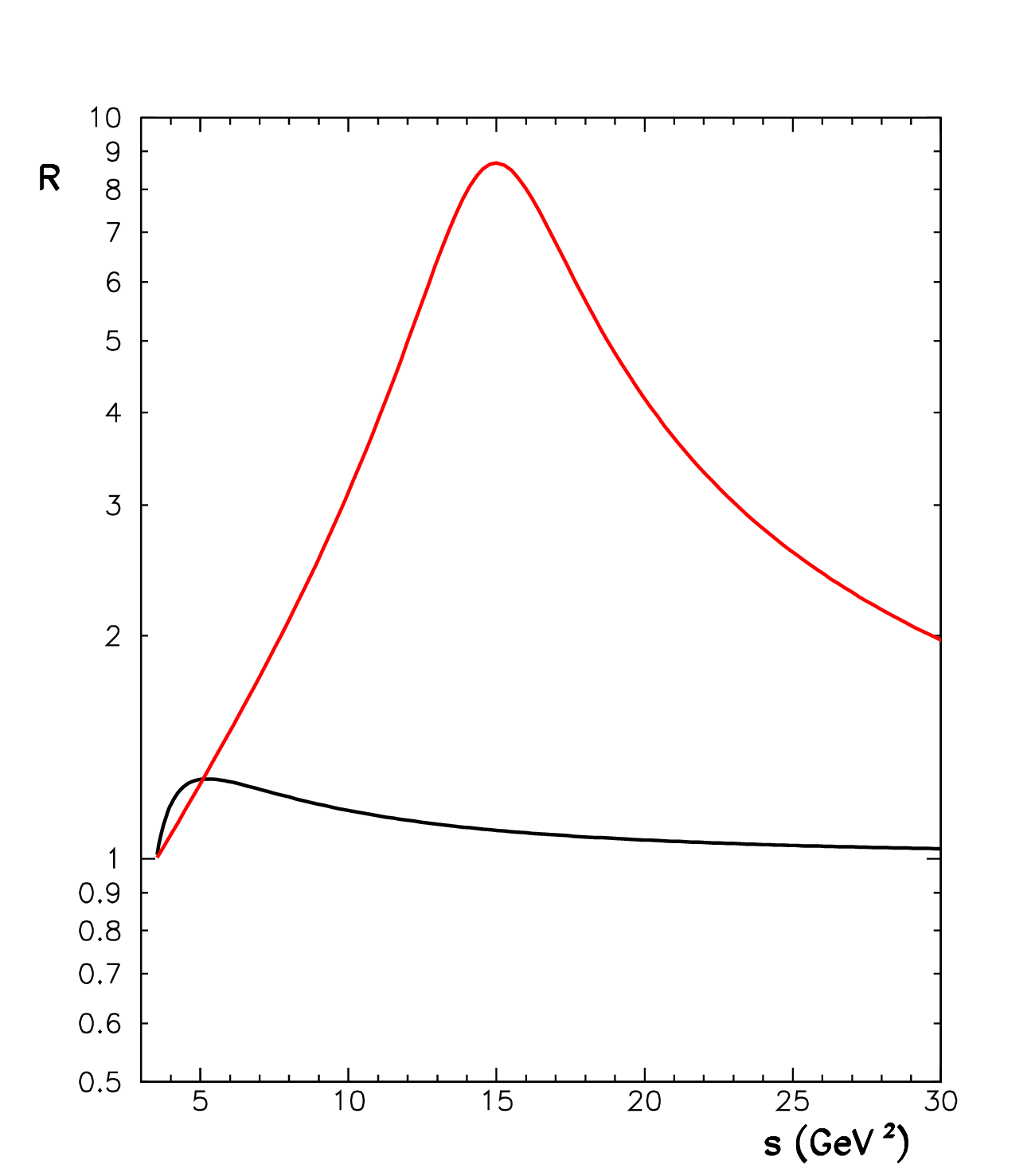}
}
\caption{Ratio $R$ versus $s$: model 1 (black line) and model 2 (red line) }
\label{Fig: 3} 
\end{center}
\end{figure}
As it can be seen, the two models are quite different and the final conclusions
concerning the effect of the electromagnetic form factors on the 
radiative corrections should be meaningful. 
\section{QED Radiative corrections to first order}

The proton electromagnetic form factors can be extracted from the angular
 distribution of the final lepton in the Born cross section. However, 
 this distribution is altered from its
 zeroth-order shape by radiative corrections. In practice, the distorted
  distribution by radiative effects can be written as :
\begin{equation}\label{eq12}
 {\displaystyle \Big[{d\sigma \over d \Omega} \Big]_R=  
 \Big[{d\sigma \over d \Omega} \Big]_B~(1+\delta)} 
\end{equation}

\begin{figure}[H]
\resizebox{0.45\textwidth}{!}{
\includegraphics{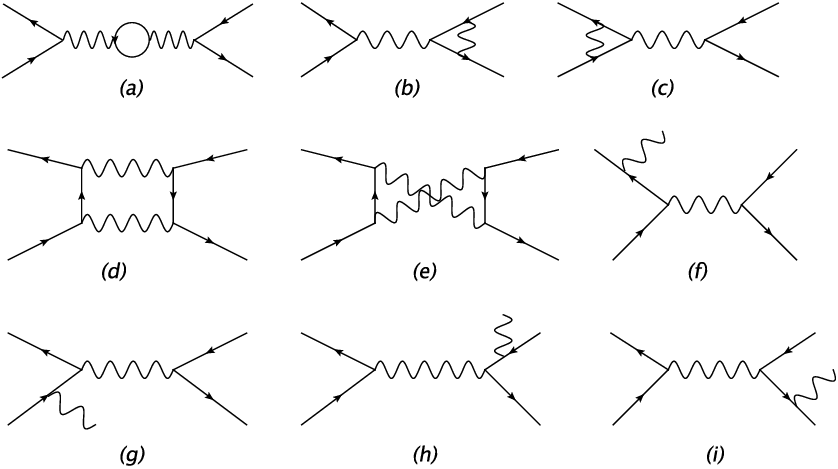}
}
\caption{Feynman diagrams for the first-order radiative correction in
 $\bar{p} p \rightarrow  e^+e^-$.}
\label{Fig: 4} 
\end{figure}

The set of diagrams contributing to the first order corrections is shown
 in Fig. 4. The virtual correction comes from the interference between the
  Born diagram (Fig. 1) and
the diagrams( (a)-(e)) of Fig. 4. The bremsstrahlung from the initial state
  (diagrams (f) and (g)) alters the effective center of mass energy and
   significantly changes
 the kinematics of the final lepton pair. And finally, the photon emission
  from the final state  is represented by the diagrams (h) and (i).
 Only the bremsstrahlung corrections lead to infrared singularities. These
  singularities are cancelled order by order by virtual corrections. 
  We adopt the standard treatment of the
 bremsstrahlung, separating the soft photon contribution with the emitted
  photon energy up to an infrared cut-off parameter $\omega$ where the soft
   photon approximation holds, 
 and the hard photon contribution from
 $\omega$ up to an experimental cut depending on the energy resolution of 
 the detector. This separation is somewhat arbitrary, 
 so we have checked that the total radiative correction (Virtual+soft+hard)
  does not depend on this infrared cut-off.\\ 
The virtual and real photon corrections are achieved by the factorization
 of the cross section in eq.(12)  with
\begin{equation}
\delta=\delta_{SV}+\delta_H
\end{equation}

We write down the soft and virtual correction together to remove the
 infrared singularities. The remaining term $\delta_{SV}$ (soft+virtual)
  is now finite. The hard photon contribution 
$\delta_H$ depends, of course, on the energy resolution of the detector.
 The full simulations in a realistic configuration of the detector will
  allow the determination of
the experimental cut on the maximum energy of the real photon emitted or
 preferably, on the invariant mass spectrum of the final lepton pair.\\

Below we present the details of our investigation of the different 
contributions to the radiative corrections in $\bar{p} p \rightarrow e^+e^-$.
\subsection{ Virtual and soft photon contributions}

 The detailed expressions of all the terms included
 in $ \delta_{SV} $  are given without any kinematical 
 approximation, which allows to use them for any value of the anti-proton
 kinetic energy and for any angle of the positron. 
\subsubsection{Virtual correction:}
As introduced at the beginning of this section, we have: 
\begin{eqnarray}\label{eq14}
\Big[ \frac{d \sigma}{d \Omega}  \Big] _{BV} =
\Big[ \frac{d \sigma}{d \Omega}\Big]_{B} 
\ \big\{ \ 1 + \delta_{V}\ \big\}
\end{eqnarray}
with $\delta_{V} $, the virtual correction
correspondig to the Feynman diagrams
$(a),(b),(c),(d),(e)$ of Fig.4 is given by:
\begin{eqnarray}\label{eq15}
\delta_{V} = \delta _{vac} + \delta ^e_{vertex} + \delta ^p_{vertex}+ \delta _{box}
\end{eqnarray}

For the vacuum polarisation contribution, the loop in diagram (a) of Fig.4
 includes the electron loop and the muon-loop [23]. 
 
\begin{eqnarray}\label{eq16}
\delta _{vac} = \frac{\alpha}{\pi} 
\Big[
      2 \,  \big(    \Pi _{e ^+ e ^- } + 
      \Pi _{\mu ^+ \mu ^- }   \big)
\Big]
\end{eqnarray}
\begin{eqnarray}\label{eq17}
\Pi _{e ^+ e ^- } = \frac{1}{3} \, \big( L_e - \frac{5}{3} \big)~;
~L_e = \ln \frac{s}{m^2}
\end{eqnarray}
\begin{eqnarray}\label{eq18}
\Pi _{\mu ^+ \mu ^-} = 
          -\frac{8}{9} + \frac{\beta ^2 _\mu}{3} 
	  + \beta  _\mu 
	       \Big( \frac{1}{2} - \frac{1}{6} \beta ^2 _\mu \Big) L_\mu
\end{eqnarray}
\begin{eqnarray}\label{eq19}
 \beta _\mu = \sqrt{1 - \frac{4 \, M^2 _\mu}{s}}
 \hspace{10mm}
 L_\mu = \ln \frac{1+\beta _\mu}{1- \beta _\mu}
\end{eqnarray}
where $  m $ and $ M_\mu $ are respectively
the electron and the muon mass. The hadronic loop is dominated by the
charged pion pair as hadron state. The form factor $ F_\pi (s) $, needed
in this contribution, will reduce strongly this pion loop contribution.
In the energy range of interest ($ s> 5 GeV^2 $) the pion loop term has 
therefore been removed from the considerations.

At the lepton vertex, the expression of  $\delta ^e_{vertex}$ was
 derived a long time ago by the authors of [24]. To deal with the infrared
  divergent term, we consider the extra virtual photon of the process
   of Fig. 4 with a mass $\lambda$.

\begin{eqnarray}\label{eq20}
&&
\delta ^e_{vertex} = \frac{\alpha}{\pi} \,
     \bigg[
              \ 2 \, ( L_e -1 ) - \frac{1}{2} L_e - \frac{1}{2} L ^2 _e 
                 + \frac{2 \, \pi ^2}{3}
\nonumber \\[1mm]
&&	\hspace{20mm}		  
		 +  2 \, ( 1 - L_e) \, \ln \frac{m}{\lambda}
            \   \bigg]		 
\end{eqnarray}

The expressions for $ \delta ^p_{vertex}  $
and $ \delta _{box}$  are derived in the  reference [18]. They assume the
 proton to be a pointlike particle. We give their result
 in this article for completeness in
 Appendix A. The validity of the pointlike 
approximation will be discussed in the next section.

\subsubsection{Real photon emission and soft photon contribution}

The real photon emission comes from the reaction
$\bar{p} p \rightarrow  \gamma e^+ e^- $.
The exact calculation of the cross section of this reaction including the electromagnetic form
factors $ F_1 $ and $F_2$ is given in the section \ref{hardphoton} and in the appendix C.1. 
The well known infrared singularity, where the energy $k_0 $ of the
 emitted photon goes to zero,   is under control by considering the soft photon
contribution. The divergent terms in  $     \delta ^e_{vertex} + \delta ^p_{vertex} + \delta _{box}  $ have to be cancelled by the divergent parts of the soft photon
 contributions. In the appendix C.2, we show analytically that the ratio between
 the soft photon contribution and the Born cross section
is independent of the model used for the form factors.\\
 Within this  soft photon approximation,  the corrected cross section is
 related to the Born cross section
using (see Appendix C.2)
\begin{eqnarray}\label{eq21}
\Big[ \frac{d \sigma}{d \Omega} 
\Big]_{\soft} =
\Big[ \frac{d \sigma}{d \Omega} \Big]_{\mathrm{B}} \ \delta_\soft
\end{eqnarray}
with 
\begin{eqnarray}\label{eq22}
\delta_\soft=-\frac{\alpha}{2 \,  \pi ^2} \
I_{\soft}
\end{eqnarray}
and
\begin{eqnarray}\label{eq23}
&&
 I_{\soft} = 
 \nonumber \\[1mm]
 && \hspace{-8mm} \hugeint _{\! \! \! 0}  ^{\omega '}
 \hspace{-2mm}
 \bigg(
     \frac{p^{+}}{k.p^{+}} - \frac{p^{-}}{k.p^{-}}
   + \frac{q^{-}}{k.q^{-}} - \frac{q^{+}}{k.q^{+}}
  \, \bigg)^2  \ \frac{d^3 \bm{k}}{2 k_0}
\end{eqnarray}
$ {\omega '} = \sqrt{\omega ^2 - \lambda ^2}$. ~$\lambda$ is again the virtual
 mass of the photon ($\lambda \rightarrow 0$).
The term $I_{\soft} $ is expanded as follows:
\begin{eqnarray}\label{eq24}
 I_{\soft} =  I_{\soft} ^{\, \mathrm{e}}
  + I_{\soft} ^{\, \mathrm{p}} +
  I_{\soft} ^{\, \mathrm{ep}}
\end{eqnarray}
\begin{eqnarray}\label{eq25}
I_{\soft} ^{\, \mathrm{e}} =
\hugeint _{\! \! \! 0}  ^{\omega '}
 \bigg(
    \frac{q^{-}}{k.q^{-}} - \frac{q^{+}}{k.q^{+}}
  \, \bigg)^2  \ \frac{d^3 \bm{k}}{2 k_0}
\end{eqnarray}
gives the contribution of the soft photon emitted by
the leptons (diagrams (h) and (i) of Fig.4).

\begin{eqnarray}\label{eq26}
I_{\soft} ^{\, \mathrm{p}} =
\hugeint _{\! \! \! 0}  ^{\omega '}
 \bigg(
     \frac{p^{+}}{k.p^{+}} - \frac{p^{-}}{k.p^{-}}
  \, \bigg)^2  \ \frac{d^3 \bm{k}}{2 k_0}
\end{eqnarray}
is the same quantity for the soft photon emitted by
the proton or the antiproton (diagrams (f) and (g) of Fig.4).

\begin{eqnarray}\label{eq27}
&& 
I_{\soft} ^{\, \mathrm{ep}} =
\nonumber \\[1mm] 
&&
\hspace{-5mm}
2 \hspace{-3mm}
\hugeint _{\! \! \! 0}  ^{\omega '} \hspace{-2mm}
 \bigg(
     \frac{p^{+}}{k.p^{+}} - \frac{p^{-}}{k.p^{-}}
 \, \bigg)  
 \bigg(    
    \frac{q^{-}}{k.q^{-}} - \frac{q^{+}}{k.q^{+}}
  \, \bigg)  \ \frac{d^3 \bm{k}}{2 k_0}
\end{eqnarray}
is the contribution of the interference between
the initial and final radiation diagrams
(diagrams (f)(h); (f)(i); (g)(h); (g)(i) of Fig.4).

The expansion (\ref{eq24}) in terms of radiative corrections reads:
\begin{eqnarray}\label{eq28}
\delta_\soft = \delta^{e}_\soft + \delta^{p}_\soft  + \delta^{ep}_\soft
\end{eqnarray}

\renewcommand{\betac}{\ensuremath{\betae}}     
\renewcommand{\etasurvl}{\ensuremath{\frac{2}{ s\, \betac }  }}
\renewcommand{\pietasurvl}{\ensuremath{\frac{2 \pi }{ s\, \betac }  }}
\renewcommand{\yzero}{\ensuremath{\frac{(1+\betac)^2}{(1-\betac)^2 }}}
\renewcommand{\yun}{\ensuremath{\frac{1-\betac}{1+\betac}}}
\renewcommand{\zun}{\ensuremath{\frac{1-\betac}{1+\betac} }}
\renewcommand{\ydeux}{\ensuremath{-\frac{2 \, \betac}{1-\betac}    }}
\renewcommand{\zdeux}{\ensuremath{0}}
\renewcommand{\ytrois}{\ensuremath{0}}
\renewcommand{\ztrois}{\ensuremath{\frac{ 2 \, \betac}{1+\betac}  }}     

The electron term in the contribution of the soft photon emitted at the lepton vertex 
can be written as 
\begin{eqnarray}\label{eq29}
&&
I_{\soft} ^{\, \mathrm{e}} =
     m ^2 \ \mathcal{I}_{q^- \, q^-} 
  +  m ^2 \ \mathcal{I}_{q^+ \, q^+}
\nonumber \\[1mm]
&&  \hspace{10mm} 
  - 2 \, q^- \cdot q^+ \ \frac{\mathcal{L}_{q^- \, q^+}}{2}
\end{eqnarray}
with
\begin{eqnarray} 
&&    
 \betac ^2   = 1-\frac{4m^2}{s} 
 \label{eq30}   \\[1mm]
 && 
    2 \, q^- \cdot q^+ = s - 2 \, m^2 = \frac{s (1+ \betac ^2)}{2}
 \label{eq31}   
\end{eqnarray}
Performing the calculation of the two first diagonal terms 
we get :
\begin{eqnarray}\label{eq32}
\hspace*{-5mm}
     m ^2 \ \mathcal{I}_{q^- \, q^-} = m ^2 \ \mathcal{I}_{q^+ \, q^+}
  = \pi \  \Big[ 2\ln \frac{ 2 \omega  }{\lambda } - \ln \frac{ s }{ m^2} \Big]
\end{eqnarray}              
Using our metric \`a la Bjorken and Drell, the calculation of the non-diagonal 
term is derived from the 't Hooft and Veltman method [25].  We give more
 details of this transposition in appendix B.\\

We need to separate the finite term (finite) from the infrared singularity term (div) 
depending on $\lambda$.    
\begin{eqnarray}\label{eq33}     
\frac{\mathcal{L}_{q^- \, q^+} ({\rm div})}{2} =    
   \ \pietasurvl 
  \ln \yzero \ \ln \frac{2\omega}{\lambda} 
\end{eqnarray}            
\begin{eqnarray}\label{eq34} 
&& 
\frac{\mathcal{L}_{q^- \, q^+} ({\rm finite})}{2} 
= 
\nonumber \\[1mm]
&&
  \pietasurvl \
     \bigg[ \    Sp \, \big( \ydeux\big )  - Sp \, \big( \ztrois \big)	    
     \ \bigg]  
\end{eqnarray}                    
\begin{eqnarray}\label{eq35}  
&&    
I_{\soft} ^{\, \mathrm{e}} = 
\nonumber \\[1mm]
&& \hspace{-5mm}
- 2 \, \pi \  \bigg\{
 \Big[ \  \frac{1+\betac^2}{2\, \betac} \ \ln \yzero \, -2 \  \Big] 
\  \ln  \frac{2\omega}{\lambda} +  \ln \frac{s}{m^2}
\nonumber \\[1mm] 
&&  \hspace{7mm}
+  \frac{ (1+ \betac ^2)}{2 \, \betac}  
 \Big[Sp \, \big( \ydeux\big )  - Sp \, \big( \ztrois \big) \Big]
  \ \bigg\} 
  \nonumber \\    
\end{eqnarray} 
Now for the soft photon emitted at the hadron vertex, 
 $I_{\soft} ^{\, \mathrm{p}}$ is derived from 
     $I_{\soft} ^{\, \mathrm{e}}$, replacing
 $\beta_e$ by $\beta_p$:
\renewcommand{\betac}{\ensuremath{\betap}}     
\renewcommand{\etasurvl}{\ensuremath{\frac{2}{ s\, \betac }  }}
\renewcommand{\pietasurvl}{\ensuremath{\frac{2 \pi }{ s\, \betac }  }}
\renewcommand{\yzero}{\ensuremath{\frac{(1+\betac)^2}{(1-\betac)^2 }}}
\renewcommand{\yun}{\ensuremath{\frac{1-\betac}{1+\betac}}}
\renewcommand{\zun}{\ensuremath{\frac{1-\betac}{1+\betac} }}
\renewcommand{\ydeux}{\ensuremath{-\frac{2 \, \betac}{1-\betac}    }}
\renewcommand{\zdeux}{\ensuremath{0}}
\renewcommand{\ytrois}{\ensuremath{0}}
\renewcommand{\ztrois}{\ensuremath{\frac{ 2 \, \betac}{1+\betac}  }}     
      
\begin{eqnarray}\label{eq36}
&&      
I_{\soft} ^{\, \mathrm{p}} =
\nonumber \\[1mm]
&& \hspace{-5mm}
 -2 \, \pi \  \bigg\{
\ \Big[  \frac{1+\betac^2}{2\, \betac} \ \ln \yzero \, -2 \  \Big] 
\  \ln  \frac{2\omega}{\lambda} + \ln\frac{s}{M^2}
\nonumber \\[1mm] 
&&  \hspace{7mm}
+  \frac{ (1+ \betac ^2)}{2 \, \betac}  
 \Big[Sp \, \big( \ydeux\big )  - Sp \, \big( \ztrois \big) \Big]
  \ \bigg\}  
  \nonumber \\   
\end{eqnarray} 

Here again, we separate the interference term  between the soft photon 
contribution from the lepton and hadron vertex into an
infrared divergent term depending on the mass $\lambda$ and a finite term:
\begin{eqnarray}\label{eq37}
I_{\soft}^{\, \mathrm{ep}} ({\rm div})
 = 4 \, \pi
       \ln \frac{M^2 -t }{M^2 -u } \ \ln \frac{M^2}{\lambda ^2}
\end{eqnarray}
\begin{eqnarray}\label{eq38}
&&
I_{\soft} ^{\, \mathrm{ep}}(\mathrm{finite}) =
\nonumber \\[1mm]
&& \hspace{-3mm}
 \ 4  \pi \,
\bigg\{ \hspace{0mm}
      2  \, \ln \frac{M^2 -t }{M^2 -u } \ \ln  \frac{2 \omega}{M} 
\nonumber \\[2mm]
&&  \hspace{4.5mm}
 + Sp \, \Big( 1 + \frac{( 1 + \betap) \,  s \, t}{2 \, M^4} \Big)
 - Sp \, \Big( 1 + \frac{( 1 + \betap) \,  s \, u}{2 \, M^4} \Big)
\nonumber \\[2mm] 
&&  \hspace{4.5mm}
 +
   Sp \, \Big(1 +   \frac{ s \, t }{(M^2 - t)^2} \Big)
 - Sp \, \Big(1 +   \frac{ s \, u }{(M^2 - u)^2} \Big) 
\nonumber \\[2mm] 
&&  \hspace{4.5mm}
 +
   Sp \, \Big(1 +  \frac{( 1 - \betap) \,  s \, t}{2 \, M^4} \Big)
 - Sp \, \Big(1 +  \frac{( 1 - \betap) \,  s \, u}{2 \, M^4} \Big)  
  \bigg\}
 \nonumber \\
\end{eqnarray}
In eqs.(34-38),
$Sp(x)$ is the dilogarithm or Spence's function defined as :

$$Sp(x)=- \int_0^x {\ln(1-t) \over t} dt$$

One can check that the infrared terms depending on $\lambda$ disappear
 when we sum up the contributions from the virtual and soft photon
  corrections.\\
We show in the table 1 and in the table 2, the soft and virtual corrections 
with final and initial state radiations as a function of the positron
angle in the center of mass system. The left columns represent the
individual contributions $ \delta_{vac} $, $ \delta^e_{vertex}$ and
$ \delta ^e_\soft $  which are usually calculated. The other 
contributions are  $\delta ^p _{vertex} $,  $ \delta^p _\soft $,
$  \delta _{box}    $ and the interference contribution  $
\delta^{ep}_\soft$ of the emitted soft photons. The right
columns give the quantities  
$$\delta ^e_{SV}  = \delta_{vac} + 
 \delta^e_{vertex} + \delta ^e_\soft$$ and the total
 contribution 
 $$  \delta ^{ep}_{SV} = \delta ^e_{SV}  + \delta ^p _{vertex} 
 + \delta^p _\soft + \delta _{box} +
 \delta^{ep}_\soft    $$
The upper limit of the energy of the soft photon, separating the soft 
and hard photon contributions is denoted by $\omega$.
Its value has been chosen such as the ratio of this
energy to the energy of the positron in the center
of mass, when no photon is emitted, is $\simeq 1 \%$.

\begin{table*}[b]
\begin{center}
\caption{\small Soft and virtual corrections : $ s = 5.4 $ GeV$^2$ , 
 $ \ \omega = 12 $ \ MeV, \ \ $\omega /
 E_{e^+} \approx 1 \% $
}
\label{tab01}
\begin{tabular}{|c|rrrrrrr|rr|}
\hline 
\raisebox{0pt}[13pt][7pt]{$\theta _{e^+} \, (deg.) $} 
&  $ \delta_{vac}     $ 
&  $ \delta ^e_{vertex}     $ & $\delta ^e_\soft  \ \   $ 
&    $ \delta ^p _{vertex}    $ & $ \delta^p _\soft  \ \  $ 
&  $ \delta _{box}   \ \    $ & $ \delta^{ep}_\soft \  $ 
&  $ \delta ^e_{SV}  \  \  \     $ & $ \delta ^{ep}_{SV} \    $     \\
\hline \hline
$  30. \ \ \   $& $ 0.0305 $ & $ -0.2602 $ & $ -0.0147 $ & $   \ \ 0.0160$ & $ -0.0127 $ & $-0.0033 $ & $ -0.0697 $ & $ -0.2444 $& $ \  -0.3108 $ \\
$  60. \ \ \   $ & $ 0.0305$ & $ -0.2602 $ & $ -0.0147 $ & $  0.0147$ & $-0.0127 $ & $-0.0020 $ & $ -0.0365 $ & $ -0.2444 $ & $ -0.2789 $ \\
$  90. \ \ \   $ & $ 0.0305 $ & $ -0.2602 $ & $ -0.0147 $ & $  0.0132$ & $-0.0127 $ & $ 0.0000 $ & $  0.0000 $ & $ -0.2444 $ & $ -0.2439 $ \\ 
$ 120. \ \ \ \ $ & $ 0.0305 $ & $ -0.2602 $ & $ -0.0147 $ & $  0.0147$ & $-0.0127 $ & $ 0.0020 $ & $  0.0365 $ & $ -0.2444 $ & $ -0.2059 $ \\   
$ 150. \ \ \ \ $ & $ 0.0305$ & $ -0.2602 $ & $ -0.0147 $ & $  0.0160$ &
$-0.0127 $ & $ 0.0033 $ & $  0.0697 $ & $ -0.244 4$ & $ -0.1713 $ \\     
\hline 
\end{tabular}
\end{center}
\end{table*}

\begin{table*}[b]
\begin{center}
\caption{\small Soft and virtual corrections : $ s = 12.9 $ GeV$^2$ ,  $ \ \omega = 18 $ \, MeV, \ \ $\omega / E_{e^+} \approx 1 \% $
}
\label{tab02}
\begin{tabular}{|c|rrrrrrr|rr|}
\hline 
\raisebox{0pt}[13pt][7pt]{$\theta _{e^+} \, (deg.) $}  
&  $ \delta_{vac}     $ 
&  $ \delta ^e_{vertex}     $ & $ \delta ^e_\soft   \ \   $ 
&  $ \delta ^p _{vertex}    $ & $ \delta ^p_\soft  \ \  $ 
&  $ \delta _{box}   \ \    $ & $ \delta^{ep}_\soft \  $ 
&  $ \delta ^e_{SV}  \  \  \     $ & $ \delta ^{ep}_{SV} \    $     \\
\hline \hline
$  30. \ \ \   $ & $ 0.0332 $ & $ -0.2991 $ & $ -0.0006 $ & $  \ \ 0.0113$&$ -0.0314 $ & $-0.0112 $ & $ -0.1169 $ & $ -0.2595 $ & $\  -0.3965 $ \\
$  60. \ \ \   $ & $ 0.0332 $ & $ -0.2991 $ & $ -0.0006 $ & $  0.0095$ &$-0.0314 $ & $-0.0059 $ & $ -0.0558 $ & $ -0.2595 $ & $ -0.3372 $ \\
$  90. \ \ \   $ & $ 0.0332 $ & $ -0.2991 $ & $ -0.0006 $ & $  0.0062$ &$-0.0314 $ & $ 0.0000 $ & $  0.0000 $ & $ -0.2595 $ & $ -0.2847 $ \\ 
$ 120. \ \ \ \ $ & $ 0.0332 $ & $ -0.2991 $ & $ -0.0006 $ & $  0.0095$ &$-0.0314 $ & $ 0.0059 $ & $  0.0558 $ & $ -0.2595 $ & $ -0.2256 $ \\   
$ 150. \ \ \ \ $ & $ 0.0332 $ & $ -0.2991 $ & $ -0.0006 $ & $  0.0113$ &$-0.0314 $ & $ 0.0112 $ & $  0.1169 $ & $ -0.2595 $ & $ -0.1627 $ \\     
\hline 
\end{tabular}
\end{center}
\end{table*}

\vspace{2mm}
 Let us add some comments on the radiative correction
(soft+virtual)  factor values in the two last columns of tables 1,2.
\begin{itemize}
\item  If only the final state radiation is taken into account, 
it is independent of the lepton scattering angle in the center of 
mass system. This is not the case for the initial state radiation, due mainly
to the interference between the initial and final state radiations
given by $\delta_\soft^{ep}$.
\item A remarquable feature is the asymmetry observed in the lepton angular 
distribution due to the charge-odd term. The angular distributions of the
 $e^+$ and $e^-$ are different.
 \item The contribution of the box diagrams (two-photon exchange) 
 is found to be  negligible, less than 1 $\%$.
 Recently, two-photon corrections in the $\bar{p} p \rightarrow  e^+ e^-$
 process were investigated with the hard rescattering mechanism [26].
 To be valid, the virtualities of both photons must be large in such an
 approach. The two-photon correction obtained is below the 1 $\%$ level.
 
\item  
The contribution of the box diagrams and the proton vertex, which depend on the
electromagnetic form factors $ F_1$ and $ F_2$ have only been calculated
in [18]
in the point-like approximation. As their contribution is small compared
to the sum of all the terms, we conclude that this approximation is
nevertheless good enough to give a reliable evaluation of the radiative
corrections.   
\end{itemize}
As we will see, the hard photon contributions in the next section do not alter
 these features. 

\subsection{Hard photon emission}\label{hardphoton}

We consider next the contribution from hard photon emission :
$$\bar{p} (p^-) ~+~p(p^+)  \rightarrow e^+(q^+) ~+~ e^-(q^-) + \gamma (k)$$
The invariant mass  $W$ of the ($e^+e^-$) system is defined by:
$$W^2=(q^+ + q^-)^2=  (p^- ~+~p^+ ~-k)^2$$

The models of the proton form factors -model 1 and model 2 - are introduced in our calculation 
of the  hard photon  contribution.

The amplitude is written as a sum of four amplitudes :
\begin{eqnarray}\label{eq39}
\mathcal{M} =    \mathcal{M}_1 + \mathcal{M}_2  +   \mathcal{M}_3
              +  \mathcal{M}_4
\end{eqnarray}
$\mathcal{M}_1,~\mathcal{M}_2, ~\mathcal{M}_3,  ~\mathcal{M}_4$ are 
respectively the amplitudes of the diagrams (h), (i), (f), (g) 
of Fig. 4. The amplitude $ \mathcal{M} $ is written as:
\begin{eqnarray}\label{eq40}
&&
\mathcal{M} (  \lambda , \lambda_{e^+}, \lambda_{e^-} ; 
\lambda_{\bar{p}}, \lambda_p) = 
\nonumber \\[1mm]
&&
\sum _i
\frac{ A^{\sigma} _i ( \lambda_{e^+}, \lambda_{e^-} ; 
\lambda_{\bar{p}}, \lambda_p )}{D_i} \
\varepsilon ^{*} _{\sigma} (k, \lambda)
\end{eqnarray}
where $  \varepsilon ^{*} _{\sigma} (k, \lambda) $ is the polarisation
vector of the emitted photon with the helicity $ \lambda $ and $  \lambda_{e^+},
\lambda_{e^-},  \lambda_{\bar{p}}$ and $\lambda_p)  $ are respectively the spin 
components of the posi-tron, the electron, the proton and the antiproton.

The expressions 
of $  A^{\sigma} _i ( \lambda_{e^+}, \lambda_{e^-} ; 
\lambda_{\bar{p}}, \lambda_p ) $  and $ D_i $  are given in Appendix C.
The sum over the photon helicity in $| \mathcal{M} |^2$ is given by
\begin{eqnarray}\label{eq41}
&& \hspace{5mm}\sum _{\lambda } 
| \mathcal{M} (  \lambda , \lambda_{e^+}, \lambda_{e^-} ; 
\lambda_{\bar{p}}, \lambda_p) |^2
\nonumber \\[2mm]
&& \hspace{-2.5mm} =
\sum ^{4}_{i j=1} 
\frac{A^{\sigma} _i ( \lambda_{e^+}, \lambda_{e^-} ;
 \lambda_{\bar{p}}, \lambda_p )
      A^{\sigma ^{ \prime ^{*}}  }_j ( \lambda_{e^+}, \lambda_{e^-} ;
       \lambda_{\bar{p}}, \lambda_p )}
		  {D_i \, D_j}
\nonumber \\[1mm]
&& \hspace{10mm}	  
 \ \sum_{\lambda}  \varepsilon ^{*} _{\sigma} (k, \lambda)	\
 	           \varepsilon _{\sigma '} (k, \lambda)    
\nonumber \\[2mm]
&&\hspace{-2.5mm} =		   
- \hspace{-1mm}
\sum ^{4}_{i j=1} 
\frac{A^{\sigma} _i ( \lambda_{e^+}, \lambda_{e^-} ;
 \lambda_{\bar{p}}, \lambda_p ) 
      A ^{*} _{ \sigma \, j} ( \lambda_{e^+}, \lambda_{e^-} ;
       \lambda_{\bar{p}}, \lambda_p )}
		  {D_i \, D_j}	
\nonumber \\[1mm]	   
\end{eqnarray}
We define the $ X_{ij} $ quantity by:
\begin{eqnarray}\label{eq42}
&&
X_{ij} =
\nonumber \\[1mm]
&&\hspace{-1mm}
 - \sum _{  \lambda_f} \hspace{-1mm}
A^{\sigma} _i ( \lambda_{e^+}, \lambda_{e^-} ; 
\lambda_{\bar{p}}, \lambda_p ) 
A ^{*} _{ \sigma \, j} ( \lambda_{e^+}, \lambda_{e^-} ;
 \lambda_{\bar{p}}, \lambda_p )
\nonumber \\[-3mm]
\end{eqnarray}
where  the sum over $ \lambda_f $ means that the sum is performed over all different fermion spin components
$ \lambda_{e^+}, \lambda_{e^-}$, $\lambda_{\bar{p}}, \lambda_p $.
Because of the fast variation of the propagators with the angle
of the emitted photon, it is convenient to write  the sum over the spins as follows:
\begin{eqnarray}\label{eq43}
&& \hspace{-4mm}
\sum_{\lambda , \,   \lambda_f}
| \mathcal{M} (  \lambda , \lambda_{e^+}, \lambda_{e^-} ; \lambda_{\bar{p}}, \lambda_p) |^2
=
\sum _{i j} \frac{X_{ij}}{D_i \, D_j}
\nonumber \\[-3mm]
\end{eqnarray}
In total, we have ten different terms:
\begin{eqnarray}\label{eq44}
&& \hspace{8mm} \sum_{\lambda , \, \lambda_f }
| \mathcal{M} (  \lambda , \,  \lambda_f|^2 =
\nonumber \\[1mm]
&&  \hspace{2mm}
        \frac{X_{11}}{D^2 _1} + \frac{X_{22}}{D^2 _2}
      + \frac{X_{33}}{D^2 _3} + \frac{X_{44}}{D^2 _4}
      + \frac{X_{12} + X_{21}}{D_1 \, D_2} 
\nonumber \\[1mm] 
&&  \hspace{-1mm}     
      + \frac{X_{13} + X_{31}}{D_1 \, D_3}     
      + \frac{X_{14} + X_{41}}{D_1 \, D_4} 
      + \frac{X_{23} + X_{32}}{D_2 \, D_3}      
\nonumber \\[1mm]
&&  \hspace{-1mm} 
      + \frac{X_{24} + X_{42}}{D_2 \, D_4}     
      + \frac{X_{34} + X_{43}}{D_3 \, D_4}             
\end{eqnarray}
The expressions of these terms $X_{ij}$ are written out 
in Appendix C.

The differential cross section in the center of mass system is given by:
\begin{eqnarray}\label{eq45}
&& \hspace{-10mm}
\frac{ d^5 \sigma }{d E_\gamma \, d \Omega _\gamma \, d \Omega _{e^+}}
= \frac{(\hbar c)^2}{32 (2\pi)^5} \, \frac{1}{\mid \bm{p^-}\mid \sqrt{s}}
  \, \frac{1}{4} 
  \nonumber \\[1mm]
  &&
  \sum_{ \lambda, \, \lambda _f}  J_{ac} \, E_\gamma \
| \mathcal{M} (  \lambda , \lambda_{e^+}, \lambda_{e^-} ; \lambda_{\bar{p}}, \lambda_p) |^2
\end{eqnarray}
where the jacobian $ J_{ac} $ is equal to:
\begin{eqnarray}\label{eq46}
J_{ac} = \frac{\mid \bm{q^+} \mid ^3}
              {\big| \mid \bm{q^+} \mid ^2 \, E_{e^-} 
	        + E_{e^+} ( \mid \bm{q^+} \mid ^2 + \bm{k}\cdot \bm{q^+} )
	       \big|}
\end{eqnarray}
This differential cross section can then be expressed as:
\begin{eqnarray}\label{eq47}
\frac{ d^5 \sigma }{d E_\gamma \, d \Omega _\gamma \, d \Omega _{e^+}}
= \sum _{ 1 \le i,j \le 4 }
\bigg[
\frac{ d^5 \sigma }{d E_\gamma \, d \Omega _\gamma \, d \Omega _{e^+}}
\bigg]_{i j}
\end{eqnarray}
\section{Total Radiative corrections and \\  Numerical Results.}
In this section, we study the radiative corrections which have to be applied
to the $ e^+ e^- $ invariant mass spectrum. These corrections depend of the
experimental cut $ W_{max} $. In the center of mass, the relation between this 
invariant mass and the photon energy is simple:
\begin{eqnarray}\label{eq48}
 s -2 E_\gamma \, \sqrt{s}  = W^2
\end{eqnarray}
We split the cross section in a virtual+soft part and a hard photon 
part as:
\begin{eqnarray}\label{eq49}
&&
\Big[ \frac {d^2  \sigma}{d\Omega_{e^+}} \Big]_R (E  ^{max} _{\gamma})
  =  
\nonumber \\[1mm]
&&  
  \Big[ \frac {d^2  \sigma}{d\Omega_{e^+}} \Big]_B \big(1 + \delta_{SV} (\omega)
   \big)
\nonumber \\[1mm]
&&   \hspace{-5mm}
 + \hugeint _{ \! \! \! \! \omega} ^{E  ^{max} _{\gamma}} \hspace{-2mm}
    \frac{d^5 \sigma}{d E_{\gamma} \, d \Omega _{\gamma} \, d \Omega _{e ^+}} \, 
     d E_{\gamma} \, d \Omega _{\gamma}
\end{eqnarray}
We expect the cross section given by eq.(49) 
to be practically independent of the cut-off $\omega$. 
\begin{figure}[H]
\begin{center}
\resizebox{0.45\textwidth}{!}{
\includegraphics{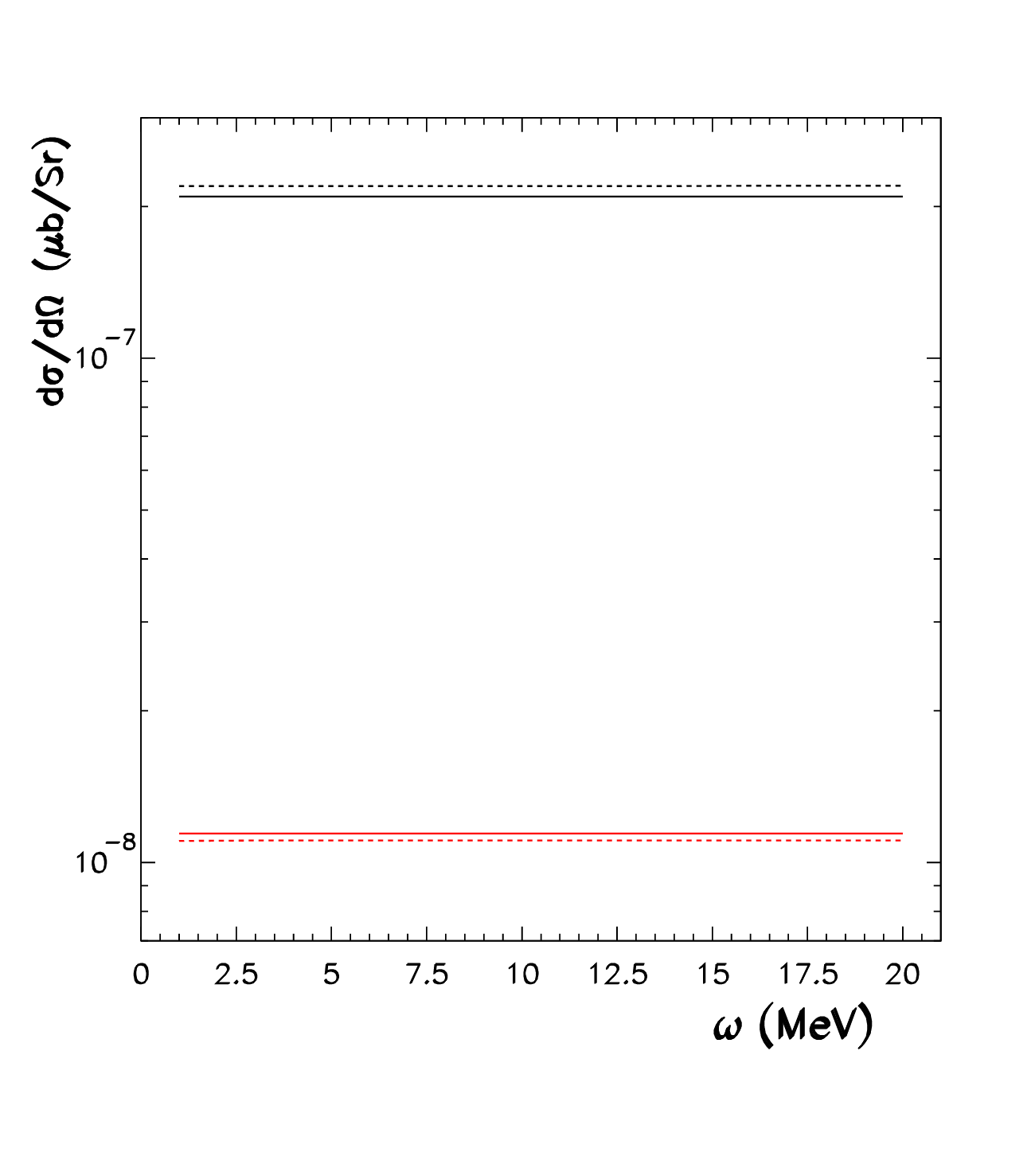}
}
\caption{Corrected cross section versus the infrared cut-off parameter
$\omega$ for  
$ s = 12.9$ \, GeV$^2 $ and $ \theta _{e^+} $= 45$^\circ$
:(model 1 : black line and model 2 : red line).
 The dotted lines correspond to the situation where only the final state
 contribution is included.}
\label{Fig: 5} 
\end{center}
\end{figure}
It is indeed what we see on fig.5, with a calulation at 
$\theta _{e^+}$ = 45 $^\circ$ and up to a photon energy $ E  ^{max} _{\gamma} $ 
of 0.4 GeV. In this
calculation, the variation  of $\omega$ is  
$ 1/1000 \leq \omega/E_{e^+} \leq 1/100$.
Within this variation of $\omega$, we do not see 
any difference in the stability with the photons emitted by the leptons alone
 or by  the 
leptons and the hadrons. Furthermore, there is no effect due to the model.
The same conclusions hold for $s = $ 5.4 GeV$^2$.

The correction factor $\delta_H$ is the ratio between the cross section with
 an extra real photon and the Born cross section, with the electromagnetic factors
 included and without the soft photon approximation.
Looking at the expression (48), it is alluring to define the 
quantities $\delta^e $ and $\delta^{ep}$ as:
\begin{eqnarray}
&&
\delta^e =  
\delta^e_{SV}(\omega) + 
  \sum _{ 1 \le i,j \le 2 }  \mathcal{R}_{ij}
\label{eq50} \\[1mm] 
&&
\delta^{ep}=\delta^{ep}_{SV}(\omega)+ 
  \sum _{1 \le i,j \le 4  }  \mathcal{R}_{ij}
\label{eq51}  
\end{eqnarray}
with
\begin{eqnarray} \label{eq52} 
 \mathcal{R}_{ij}  = \hspace{-5mm}      
  \hugeint _{ \! \! \! \! \omega} ^{E  ^{max} _{\gamma}} \hspace{-5mm}
 \bigg[    \frac{d^5 \sigma}{d E_{\gamma} \, d \Omega _{\gamma} \,
  d \Omega _{e ^+}} \bigg]_{i j}  \hspace{-1mm} d E_{\gamma} d \Omega _{\gamma} 
  \hspace{-3mm}
      \hugeslash \hspace{-1mm} 
      \Big[ \frac {d^2  \sigma}{d\Omega_{e^+}} \Big]_B 
      \nonumber \\
\end{eqnarray}
We now give the numerical results associated with these
formula. The variation of the cross section as a function
of the invariant mass square of the $(e^+e^-)$ system is shown in Fig. 6.
\begin{figure}[H]
\begin{center}
\includegraphics*[bb = 1 50 590 660,width=6cm]{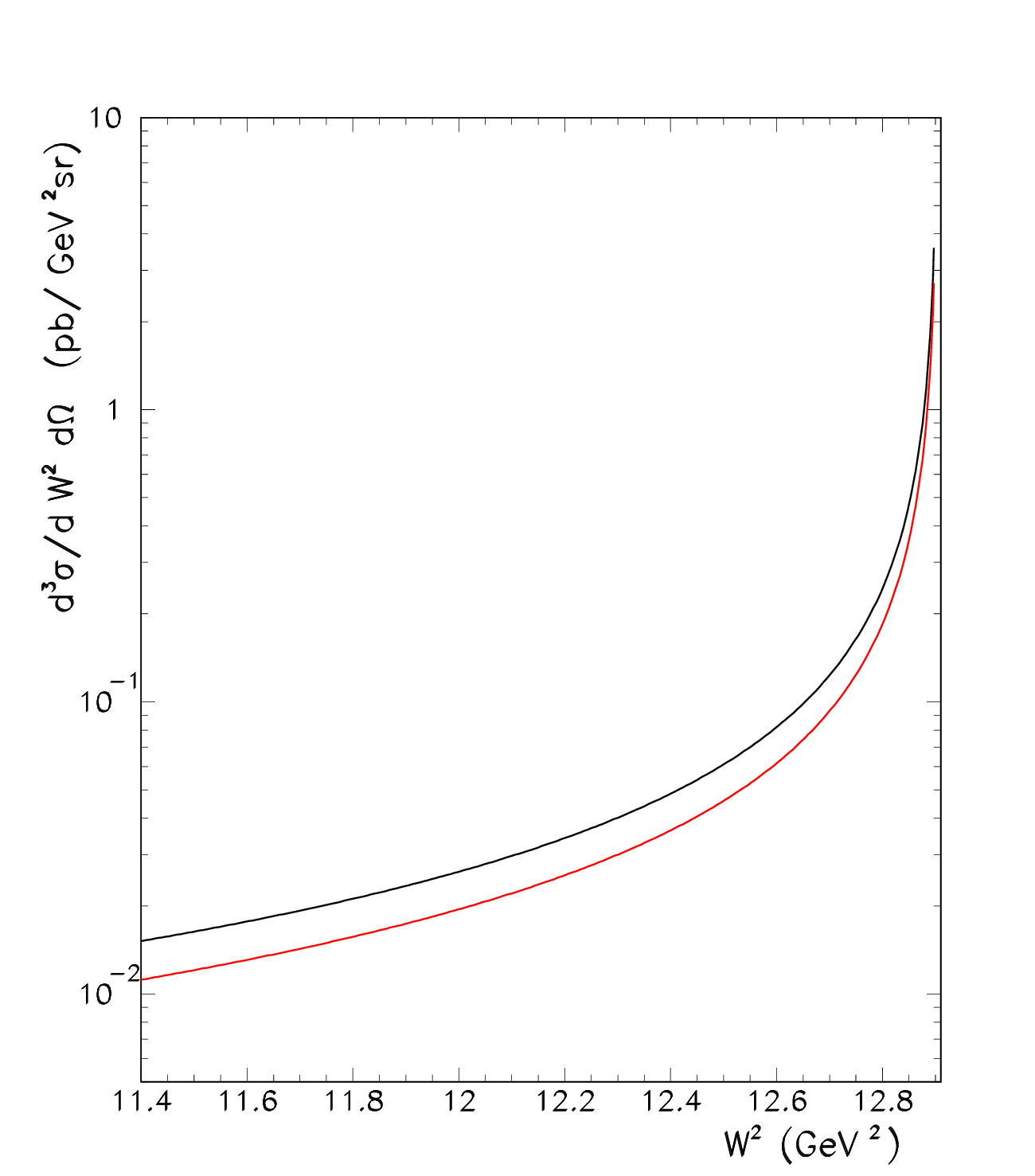}
\end{center}
\caption{(Color online) Differential cross section 
$d^2 \sigma /dW^2 d\Omega_{e^+}$  as a function of $W ^2$ 
($W$ is the invariant mass of the $(e^+e^-)$ system), 
 for   $ s = 12.9$\, GeV$^2 $,
$E_{\gamma}^{max}=100$ MeV  and $\theta_{e^+}  =30^0 $ (model 1 :
 black line and model 2 : red line) }
\label{Fig: 6} 
\end{figure}
The numerical results strongly depend on the experimental energy cut-off of the
 emitted photon.
 We have chosen to give in this article the results for a cut-off
energy $E_{\gamma}^{max} $  of 0.1 GeV. The values of the $\delta $ parameters
for some positron angles are given in Table 3 using model 2.
\begin{table}[H]
\begin{center}
\caption{\small Total radiative corrections for $ s = 5.4 \, $ GeV$^2$
  (two left columns) and for $ s = 12.9$ \, GeV$^2 $ (two right columns),
   assuming
the energy of the hard photon emission up to $100$~ MeV}
\label{tab03}
\begin{tabular}{|c|rr|rr|}
\hline 
\multicolumn{5}{|c|}{\raisebox{0pt}[12pt][7pt]{\hspace{22mm}$ s=5.4\, $ GeV$^2$ 
\hspace{10mm}$ s = 12.9$ \, GeV$^2 $}}\\
\hline
\raisebox{0pt}[12pt][7pt]{$\theta _{e^+} \, (deg.)  $} 
&  $ \delta ^e \ \ \    $ & $ \delta ^{ep} \  $  
&  $ \delta ^e  \  \       $ & $ \delta ^{ep} \    $   \\
\hline \hline
$ 30. \ \     $ & $  -0.0952   $ & $ -0.1300  $ & $ -0.1320    $ & $ -0.2191   $  \\
$ 60. \ \     $ & $  -0.0952   $ & $ -0.1096  $ & $ -0.1320    $ & $ -0.1783   $  \\
$ 90. \ \ \   $ & $  -0.0952   $ & $ -0.0878  $ & $ -0.1320    $ & $ -0.1419   $  \\ 
$ 120. \ \ \  $ & $  -0.0952   $ & $ -0.0629  $ & $ -0.1320    $ & $ -0.0986   $  \\   
$ 150. \ \ \  $ & $  -0.0952   $ & $ -0.0396  $ & $ -0.1320    $ & $ -0.0528   $  \\     
\hline 
\end{tabular}
\end{center}
\end{table}
The dependence of the radiative
correction with the positron angle $ $ is displayed in Figs. 7-8.
\begin{figure}[H]
\begin{center}
\includegraphics*[bb = 1 50 590 660,width=6cm]{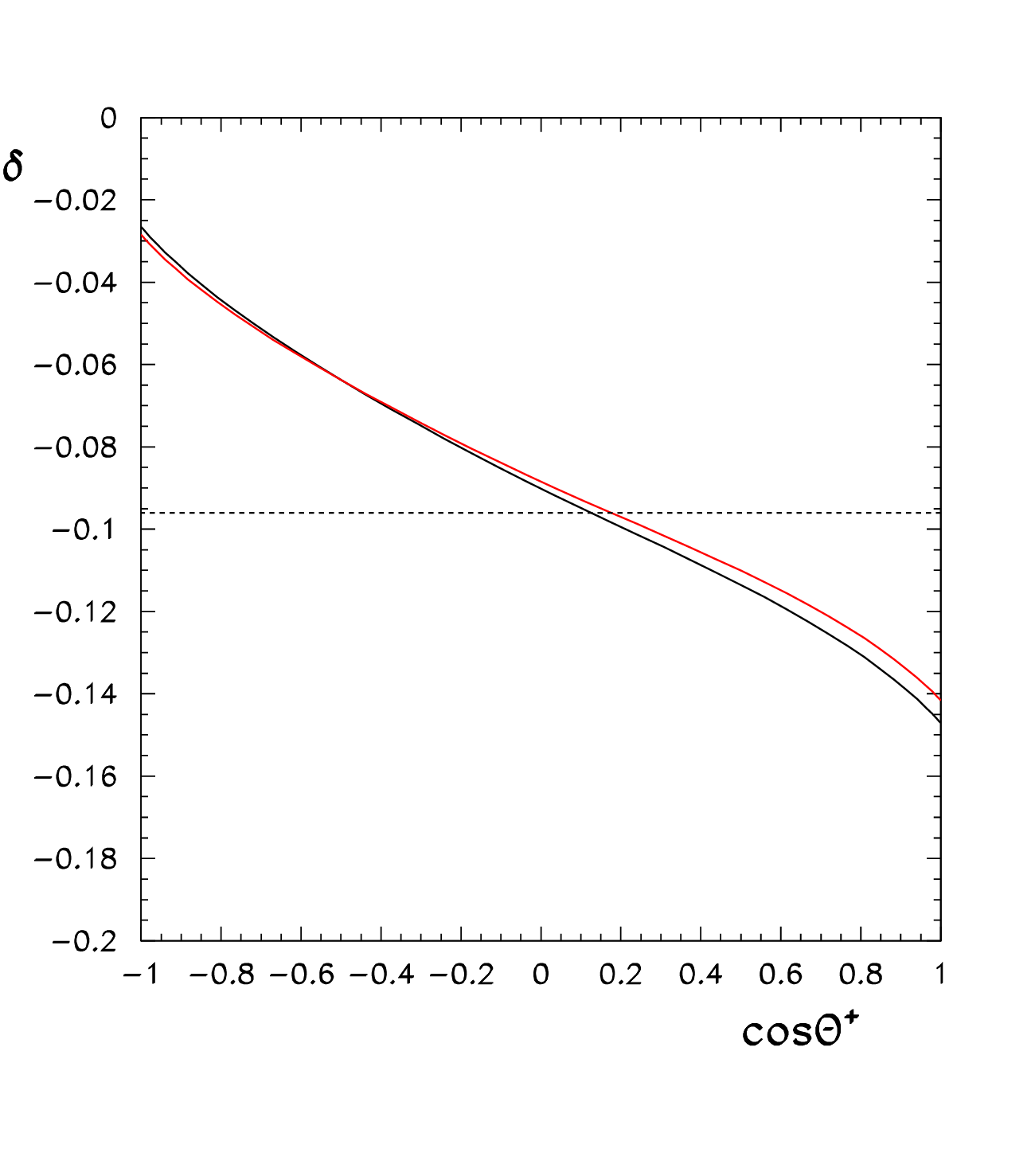}
\end{center}
\caption{(Color online) Total radiative correction  $\delta$ as a 
function of $\cos \theta _{e^+}$ in the CM frame for   $ s = 5.4 $\, GeV$^2 $
and $E_{\gamma}^{max}=100$ MeV (model 1 : black line and model 2 :
 red line). The dashed line corresponds to the situation where only the final state
radiation contributes.}
\label{Fig: 7} 
\end{figure}
\begin{figure}[H]
\begin{center}
\includegraphics*[bb = 1 50 590 660,width=6cm]{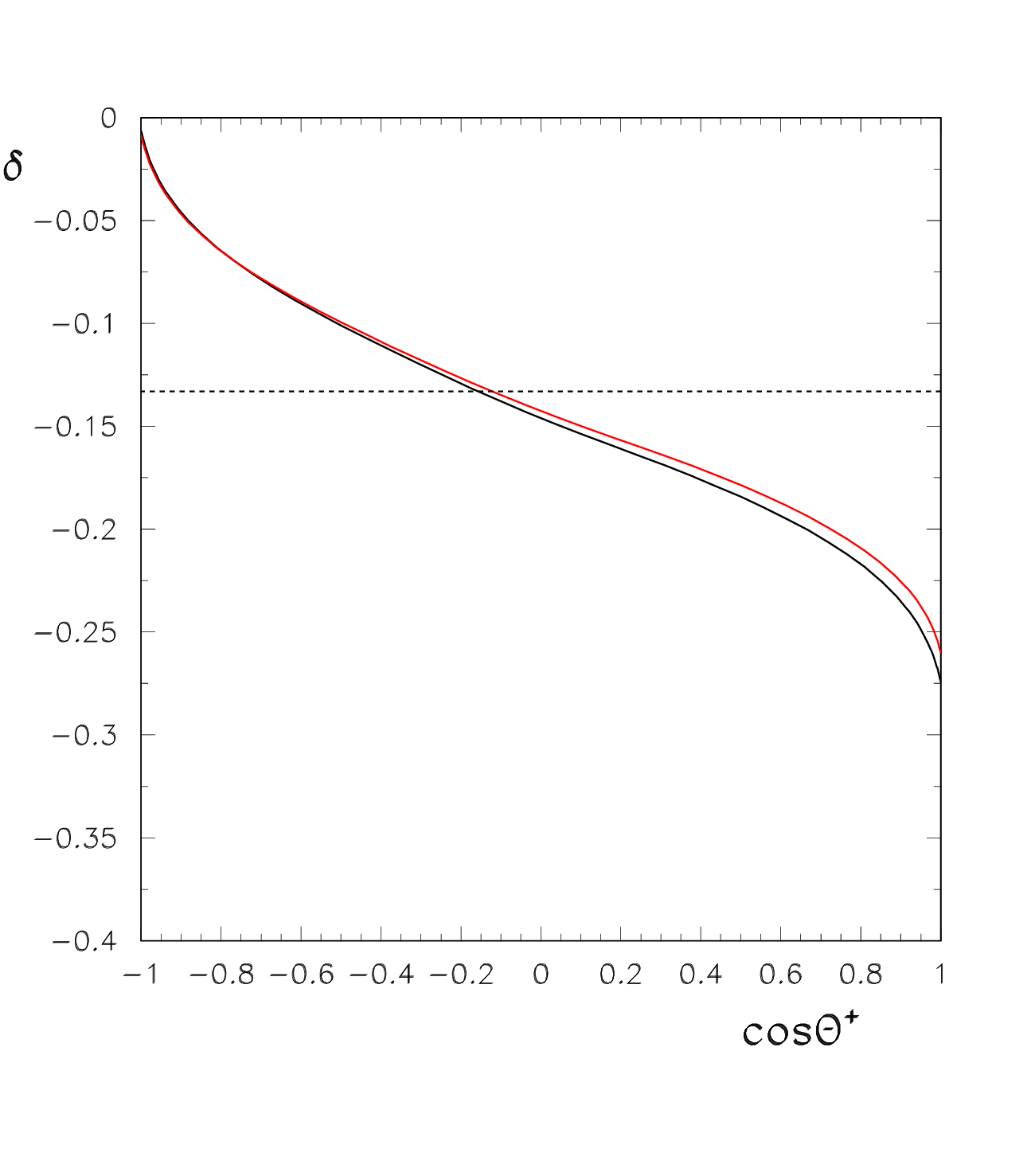}
\end{center}
\caption{(Color online) Total radiative corrections factor $\delta$ as a 
function of $\cos \theta_{e^+}$ in the CM frame for   $ s = 12.9$\, GeV$^2 $
and $E_{\gamma}^{max}=100$ MeV (model 1 : black line and model 2 : 
red line) . The dashed line corresponds to the situation where only the final state
radiation contributes.}
\label{Fig: 8} 
\end{figure}
If only the final state radiation contributes, the value of $\delta$
is independent of the positron angle in the C.M. system. In contrast,
 when the initial and final state radiations are taken into account,
 we notice large differences in the value of the correction factor
 $ \delta $ at backward and forward positron angles (Figs:7,8).
The main reason for this effect is  the interference term.
Let us emphasize the practically independent correction factor $ \delta$
with the model used for the electromagnetic form factors.

The corrected cross section is displayed for comparison with the Born
 cross section in Fig. 9. The different curves are obtained using the model 1
for the form factors. Of course the normalisation of these cross sections
 depend on the model assumptions. One can also  remark the asymmetry of the
  black line due to the charge-odd term when the the initial state radiation at 
  the hadron vertex is included. The measurement of this asymmetry term 
  included in the angular distribution seems to be a difficult task.

\begin{figure}[H]
\begin{center}
\includegraphics*[bb = 1 50 590 660,width=6cm]{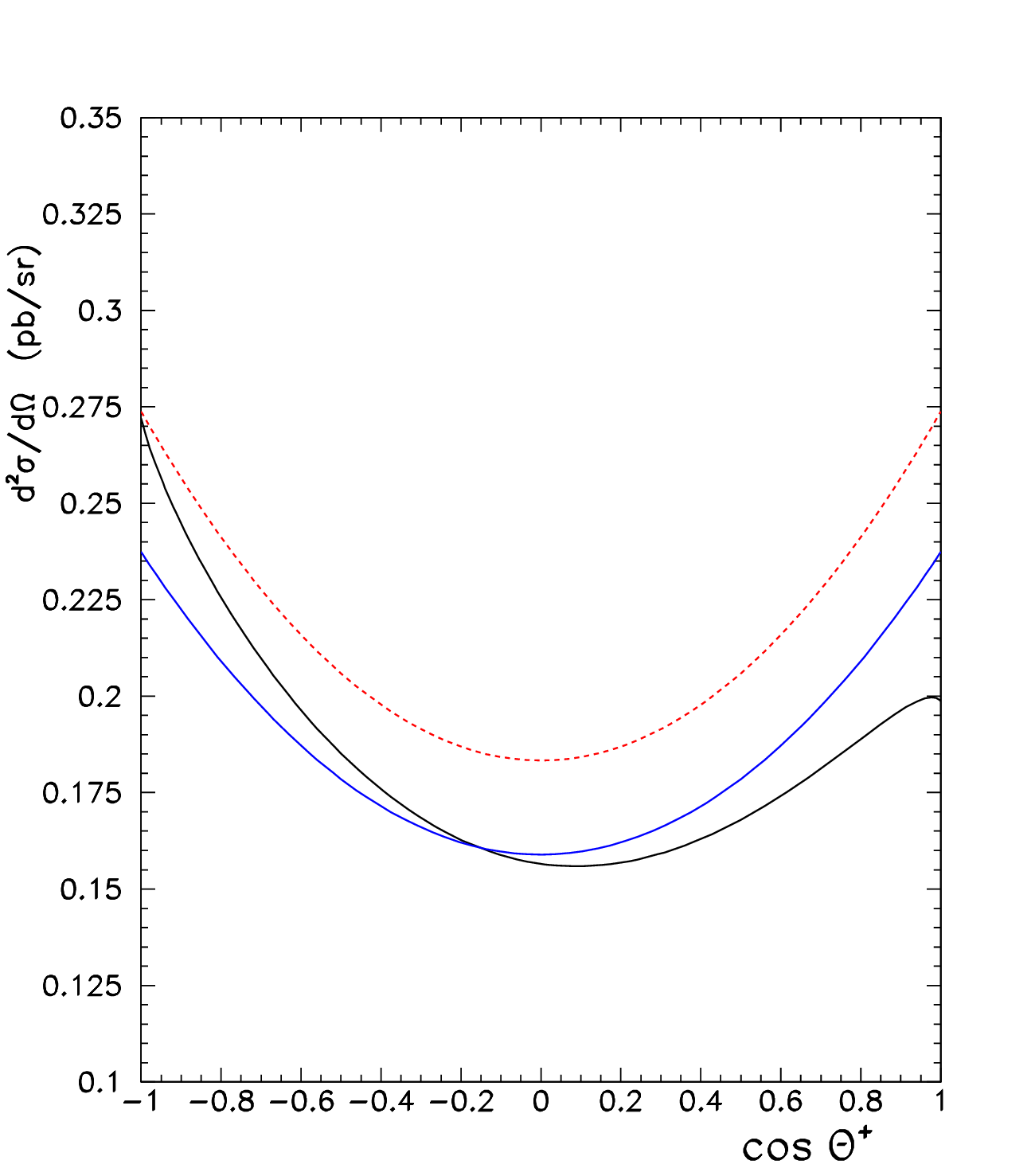}
\end{center}
\caption{(Color online)  Corrected differential cross section 
$d^2 \sigma /d\Omega_{e^+}$ as a function of $\cos \theta_{e^+}$ in the CM frame,
 for   $ s = 12.9$\, GeV$^2 $,
$E_{\gamma}^{max}=100$ MeV  (black line). The corresponding blue line 
corresponds to the situation where only the final state
radiation contributes. The red dotted line is the Born cross
  section with the assumption model 1 for the form factors. }
\label{Fig: 9} 
\end{figure}
The total radiative correction have an important symmetry that is worth to 
mention.
The values of $\delta$ are the corrections to the angular distribution of 
the positron $(e^+)$. The corresponding distribution of the electron  $(e^-)$
 is obtained by
replacing ($\theta$) by  ($\pi -\theta$) in order to respect the C-charge 
symmetry. We have checked numerically that the total radiative correction 
reads :
$$\delta ^{(e^+)}(\theta)=\delta^{(e^-)}(\pi-\theta)$$

This observation led us to define two interesting observables namely:
\begin{eqnarray}\label{eq53}
\mathcal{S}= \frac{1}{2} 
   \left(\frac{d \sigma}{d\Omega_{e^+}} 
+        \frac{d \sigma}{ d\Omega_{e^-}}\right)
\end{eqnarray}
and
\begin{eqnarray}\label{eq54}
\mathcal{A}=
\left(
{d \sigma \over d\Omega_{e^+}} -{d \sigma \over d\Omega_{e^-}}
\right)
  \Bigg /  \left(
  {d \sigma \over d\Omega_{e^+}} 
  +{d \sigma \over d\Omega_{e^-}}
  \right) \label{54}
\end{eqnarray}    
The first one contains the charge-even terms and is the corrected 
 Born cross section for form factors extraction.
 For this observable, we have
\begin{eqnarray}\label{eq55}
\frac{d\sigma}{d\Omega_e} = \Big[ \frac{d\sigma}{d\Omega_e} \Big]_B ( 1+ \delta ^{\mathcal{S}}) 
\end{eqnarray}
with
\begin{eqnarray}\label{eq56} 
\delta^{\mathcal{S}} = 
\frac{1}{2} \, \big[ 
\delta (\theta) + \delta (\pi - \theta ) \big]
\end{eqnarray}

\begin{figure}[H]
\begin{center}
\includegraphics*[bb = 1 50 590 660,width=6cm]{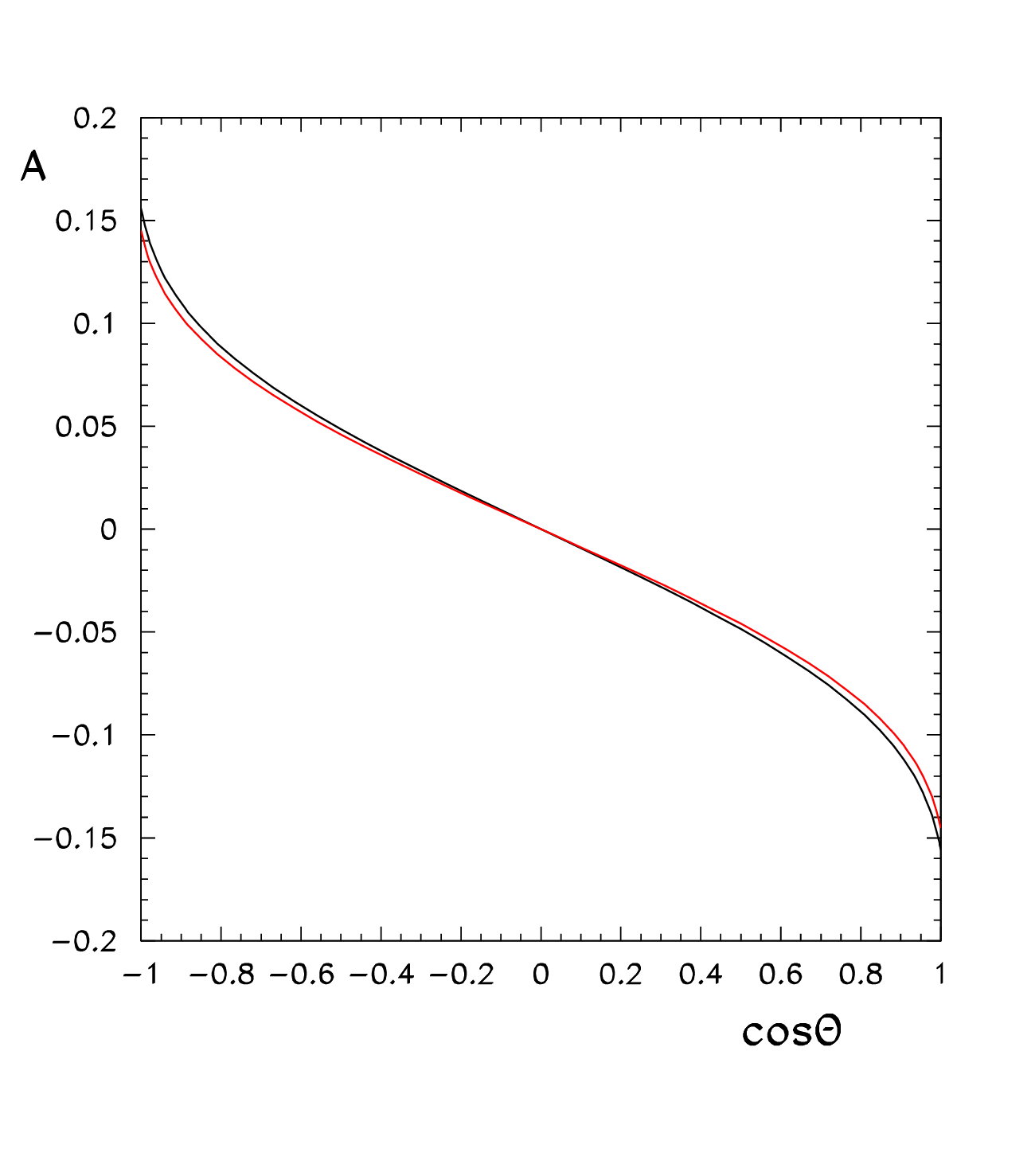}
\end{center}
\caption{(Color online) Asymmetry $\cal {A}$ versus $\cos \theta$ 
for   $ s = 12.9$\, GeV$^2 $: for $E_{\gamma}^{max}=100$ MeV
(model  1 : black line and model 2 : red line)} 
\label{Fig: 10} 
\end{figure}

 The second
one is the charge asymmetry observable due to the odd part. The value of 
this charge asymmetry term ${\cal A}$ is rather large and can be measured with
the PANDA detector (Fig. 10). The shape 
and the normalization of the observable ${\cal S}$ is model-dependent and could be used 
to discriminate between different models of the form factors.

\section{ Conclusions}

We conclude that the initial state radiation (at the hadron vertex) is not
 negligible and can be calculated and incorporated into the total radiative
  correction. If the precision of the
cross section measurement is of the order of 3-5$\%$, the total 
model independent correction
 obtained in this paper can be used to correct 
 the measured cross section, before the comparaison with the Born cross 
 section needed to discriminate between the different model assumptions
  of the proton form factors.

We have shown that the interference between the initial and final state 
radiations is the main contribution to the charge asymmetry $ {\cal A}$. 
The evaluation of this radiative correction in terms of structure functions
 and evolution equations [27]
must be performed with caution. A study of the angular distribution of the
 photon shows a non negligible part of the photons emitted outside the cone
  of angle 
$\theta_{\gamma} \leq m/E$.\\

In contrast to space-like elastic electron-proton scattering, in the 
time-like region we can consider the shape of the $e^+$ and $e^-$ angular
 distributions separately to exhibit the charge asymmetry.
The numerical result displayed in Fig. 9 shows the limit of this statement.
 We suggest the measurement of two observables namely ${\cal S}$ and 
 ${\cal A}$.\\

With the PANDA detector, an accurate calculation of the differential 
cross section of the lepton pair and of the real photon is needed
to disentangle the $ \bar{p} p \rightarrow (\gamma) e^+ e^-$
reaction from the $ \bar{p} p \rightarrow \pi ^+$ $\pi ^- $ reaction.
As the pion counting rate is huge compared to the lepton counting
rate, the angular distribution of the real photon has to be known
as precisely as possible, so the photon emission in the proton
anti-proton side can not be ignored. 
Based on the formalism described in this article, we have developed
a Monte Carlo Code.  A full simulation 
 is needed to take into account the radiative corrections together with 
efficiency and acceptance corrections.
Based on the Babar data, the model 1 for the electromagnetic form factors
allows to have a realistic calculation of the 
$  \bar{p} p \rightarrow (\gamma) e^+ e^-$ process. Contrary to
the model 2 and the point-like model, it gives the good order
of magnitude of the counting rate.

\section{Acknowledgments}

The authors would
like to thank the ORSAY/PANDA Collaboration members for constructive remarks
and constant encouragements, in particular R. Kunne for a careful reading
 of the manuscript.

\begin{appendix}

\section{ Proton vertex  and Box diagrams contribution}

The derivation given in this appendix is due to the work of the authors 
of ref[18].


For $\delta ^p_{vertex}  $ corresponding to
the Feynman graph $ (c)  $  of Fig. 4, the correction is given by:
\begin{eqnarray}\label{eq57}
\delta ^p_{vertex} = \frac{\alpha}{\pi} \, F ^p _{vertex}
\end{eqnarray}
where $ F ^p _{vertex} $ is the sum of two
terms.: 
\begin{eqnarray}\label{eq58}
F ^p _{vertex}=
2 F^{(2)} _1 + \frac{ 4 \,F^{(2)} _2}{ 2 -  \betap^2 \, \sin^2 \theta }
\end{eqnarray}
The first one contains the infrared divergence:
\begin{eqnarray}\label{eq59}
&&  \hspace{-12mm}
 F^{(2)} _1 =  
    -1  - \frac{1}{4 \, \betap} \, L_{\betap}    
\nonumber \\[1mm]  
&& 
\hspace{-2mm}
   + \frac{1+\betap ^2}{2 \, \betap}
     \bigg[ \frac{\pi ^2}{3} +L_{\betap}  +  Sp \, 
     \big( \frac{1-\betap}{1+\betap} \big)
            - \frac{1}{4} L^2 _{\betap}
\nonumber \\[1mm]  
&& 
\hspace{13mm}	     
	    - L_{\betap} \, \ln \frac{2 \, \betap}{1+\betap}
     \bigg]
\nonumber \\[1mm] 
&&  \hspace{-2mm}
   +  \Big( 1 -  \frac{1+\betap ^2}{2 \, \betap} \, L_{\betap} \Big)
       \, \ln \frac{M}{\lambda}
\end{eqnarray}
while the second term is finite:
\begin{eqnarray}\label{eq60}
 \frac{ 4 \,F^{(2)} _2}{ 2 -  \betap^2 \, \sin^2 \theta }
 = - \frac{1-\betap ^2}{\betap} \ \ 
 \frac{L_{\betap}}{2 -  \betap^2 \, \sin^2 \theta}
\end{eqnarray}
\begin{eqnarray}\label{eq61}
L_{\betap} = \ln \frac{1+\betap}{1-\betap}
\end{eqnarray}

To give the formulas related to the box
diagrams contribution, we first remind the reader the usual
expressions for the Mandelstam variables 
$s$, $t$ and $u$ of the process 
 namely
$$s=(p^-+p^+)^2~,~t=(p^- - q^+)^2~,~u=(p^-- q^-)^2$$

The two-photon correction is given by:
\begin{eqnarray}\label{eq62}
\delta_{box} 
   = \frac{ \alpha }{\pi} \,
     \frac{ s \, I(s,t,u)}{\Delta}
\end{eqnarray}
with
\begin{eqnarray}\label{eq63}
\Delta = \dborn
\end{eqnarray}

The quantity $ I(s,t,u)$ (Eq.(20) of Ref.[18]) is the sum of five terms:
\begin{eqnarray}\label{eq64}
 I(s,t,u) = \sum_{i=1}^5 I^i(s,t,u)
\end{eqnarray}
The infrared divergence is contained in  the last term:
\begin{eqnarray}\label{eq65}
\frac{ s I ^5 (s,t,u) }{\Delta} =  2 \,  L_{tu} \, L_{M \lambda} 
                 + 2 \,  L_{tu} \, L_{s} 
\end{eqnarray}
Following eq.(\ref{eq64}), we have
\begin{eqnarray}
\delta_{box} = \sum _{i=1}^5 \delta_{i \ box}
 \label{eq66}
\end{eqnarray}
\begin{eqnarray}\label{eq67}
&&
\delta_{1 \ box}= \frac{\alpha}{\pi}
\nonumber \\[1mm]
&&
 \frac{ s \, (u-t)}{\Delta} \,
  \bigg[ \Big( \frac{2 M^2}{\betap ^2} + t + u \Big) \, I_{0} 
          - \frac{\pi ^2}{6}  + \frac{1}{2} L_{\betap} ^2 
	  - \frac{L_{\betap}}{\betap ^2} 
  \bigg]
  \nonumber \\
\end{eqnarray}
with
\begin{eqnarray}\label{eq68}
&& \hspace{-2mm}
I_{0} = \frac{1}{ s  \betap}
 \bigg\{
  L_s L_{\betap} - \frac{1}{2} L_{\betap} ^2 - \frac{\pi ^2}{6}
   +   2   Sp \Big( \frac{1+\betap}{2} \Big)  
\nonumber \\[1mm]  
&& \hspace{14mm}
 - 2   Sp \Big( \frac{1-\betap}{2} \Big)
 - 2   Sp \Big(-\frac{1-\betap}{1+\betap} \Big) 
 \bigg\}
 \nonumber \\
\end{eqnarray}
and
\begin{eqnarray}\label{eq69}
\hspace*{-2mm}
\delta_{2 \ box}  =  \frac{\alpha}{\pi}
\ \frac{ s \, (2t+s)}{\Delta} \, 
  \bigg[ 
         \frac{1}{2} L_{ts} ^2 - Sp\Big( \frac{-t}{M^2 - t} \Big)
  \bigg]
\end{eqnarray}
\begin{eqnarray}\label{eq70}
\hspace*{-2mm}
\delta_{3 \ box}  = -\frac{\alpha}{\pi}
 \frac{ s \, (2u+s)}{\Delta} \, 
  \bigg[ 
         \frac{1}{2} L_{us} ^2 - Sp\Big( \frac{-u}{M^2 - u} \Big)
  \bigg]
\end{eqnarray}
\begin{eqnarray}\label{eq71}
&& \hspace{-5mm}
\delta_{4 \ box}  =  \frac{\alpha}{\pi}
 \frac{ s \, \big( ut - M^2 (s+M^2)\big)}{\Delta} 
\nonumber \\[1mm]
&& \hspace{15mm}
  \bigg[ 
         \frac{L_{ts}}{t} - \frac{L_{us}}{u} + \frac{u-t}{ut} \, L_{s}
  \bigg]  
\end{eqnarray}
\begin{eqnarray}\label{eq72}
\delta_{5 \ box}  =  \frac{\alpha}{\pi} \, \big(
2 \, L_{tu} \, L_{s}  + 2 \,  L_{tu} \, L_{M \lambda} \big)
\end{eqnarray}
\begin{eqnarray}\label{eq73}
L_{s} = \ln \frac{s}{M^2}  \hspace{10mm}
L_{tu}= \ln \frac{M^2 - t}{M^2 - u}
\end{eqnarray}
\begin{eqnarray}\label{eq74}
L_{ts}= \ln \frac{M^2 - t}{s}
  \hspace{10mm}
L_{us}= \ln \frac{M^2 - u}{s}  
\end{eqnarray}
\begin{eqnarray}\label{eq75}
L_{\betap} = \ln \frac{1+\betap}{1-\betap}
\hspace{10mm}
L_{M \lambda} = \ln \frac{ M^2}{\lambda ^2}
\end{eqnarray}

\section{'t Hooft and Veltman integrals}

We need to calculate the integral of the type :
\begin{eqnarray}\label{eq76}
\frac{\mathcal{L}_{ij}}{2} =
\hugeint _{\! \! \! 0}  ^{\omega '}
\frac{1}{ (p_i \cdot k) (p_j \cdot k)}\, \frac{d^3 \bm{k}}{2 k_0}
\end{eqnarray}
where $ {\omega '} = \sqrt{\omega ^2 - \lambda ^2}$ and  $ \omega =(k_0)_{max}
$ is the maximum energy of the soft photon.
\begin{eqnarray}
&&
 p = \eta  \, p_i
\hspace{8mm} q = p_j 
\hspace{8mm} (p-q)^2 = 0
\label{eq79} \\[1mm]
&& 
  \eta ^2 \, p_{i}^2 - 2 \eta \, p_i \cdot p_j + p_{j}^2 =0
\label{eq81}\\[1mm]  
&&
 \ell = p_0 - q_0
 \hspace{10mm}  v = \frac{ p ^2  - q ^2 }{2 \, \ell}
\label{eq82} 
\end{eqnarray}
with the condition that 
$ (\eta   p_i - p_j)_{_0} $ and $ p_{j_{0}} $ have the same sign.

The integral $\mathcal{L}_{ij} $ is the sum of a divergent part and
a finite part. Using Spence's function rather than its approximation, 
 we get:
\begin{eqnarray}\label{eq83}
\mathcal{L}_{ij}({\rm div}) = 
2\pi \,   \frac{\eta}{ v \, \ell}  \  
  \ln \frac{p ^2}{q ^2} \ \ln \frac{2\omega}{\lambda} 
\end{eqnarray}
\begin{eqnarray}\label{eq84}
&& \hspace{-1mm}
\mathcal{L}_{ij}({\rm finite}) =
\nonumber \\[1mm]
&& \hspace{2mm}
   2  \pi   \frac{\eta}{ v \, \ell}  \
     \bigg[ \
             \frac{1}{4} \ln ^2 \frac{p_0 - \modvecp  }{p_0 + \modvecp } 
          -  \frac{1}{4} \ln ^2 \frac{q_0 - \modvecq  }{q_0 + \modvecq } 
    \ \bigg]       
\nonumber \\[3mm]
&&     
 + 2\pi \frac{\eta}{ v \, \ell}  \        
     \bigg[ \
             Sp \Big( 1 - \frac{p_0 + \modvecp}{v} \Big)
	   - Sp \Big( 1 - \frac{q_0 + \modvecq}{v} \Big)
    \ \bigg]
\nonumber \\[3mm]
&&     
 + 2\pi \, \frac{\eta}{ v \, \ell} \        
     \bigg[ \
             Sp \Big( 1 - \frac{p_0 - \modvecp}{v} \Big)
	   - Sp \Big( 1 - \frac{q_0 - \modvecq}{v} \Big)
     \ \bigg]
     \nonumber \\
\end{eqnarray}
\section{Real photon emission.}

\subsection{Hard photon contribution.}

\subsubsection{Amplitudes.}

 Defining
\begin{eqnarray}
&& \hspace{-5mm}
\Gamma ^{\sigma}_{NN \gamma } (k) = F_{10} (k^2) \gamma ^{\sigma} 
- \frac{F_{20} (k^2)}{4 M }
 (\kslash \gamma ^{\sigma} - \gamma ^{\sigma} \kslash) 
 \label{eq85} \\[1mm]
 &&
 F_{10} (k^2) = 1 \hspace{10mm} F_{20} (k^2)=\kappa _p
\label{eq86} 
\end{eqnarray}
where $\kappa _p $ is the anomalous magnetic moment of the proton and 
\begin{eqnarray}\label{eq87}
 q' = \pmoins + \pplus - k = q -k
\end{eqnarray}
\begin{eqnarray}\label{eq88}
&&\hspace{-10mm}
\mathcal{M}_i ( \lambda , \lambda_{e^+}, \lambda_{e^-} ; \lambda_{\bar{p}}, 
\lambda_p ) =
\nonumber \\[1mm]
&& \hspace{-2mm}
 \frac{ A^{\sigma} _i (\lambda_{e^+}, \lambda_{e^-} ;
  \lambda_{\bar{p}}, \lambda_p )}{D_i} \  
\varepsilon ^{*} _{\sigma} (k, \lambda)
\hspace{3mm} i=1,4
\end{eqnarray}
\begin{eqnarray}\label{eq89}
&&\hspace{-3mm}
 A^{\sigma} _1 ( \lambda_{e^+}, \lambda_{e^-} ; \lambda_{\bar{p}}, \lambda_p ) 
 = \frac{i}{q^2} \, e_{p} \, e^2_{e^-} \ 
 \nonumber \\[1mm]
 &&
\Big[
\vbarpbar  \,  \Gamma _{NN \gamma }  ^{\mu} (q) \, \up 
\Big]
 \hspace{0mm}
 \Big[
 \ubarem  \,  \gamma _{\mu} \, \big( -\kslash - \qplusslash + m \big)
\,  \gamma ^{\sigma} \, \vep 
 \Big] 
\nonumber \\ 
\end{eqnarray}
\begin{eqnarray}\label{eq90}
D_1 = (k+\qplus)^2 - m^2 = 2 \, k \cdot \qplus
\end{eqnarray}
\begin{eqnarray}\label{eq91}
&&\hspace{-3mm}
 A^{\sigma} _2 ( \lambda_{e^+}, \lambda_{e^-} ; \lambda_{\bar{p}}, \lambda_p ) =
\frac{i}{q^2} \, e_{p} \, e^2_{e^-} \
 \nonumber \\[1mm]
 &&
\Big[
\vbarpbar  \,  \Gamma _{NN \gamma }  ^{\mu} (q) \, \up 
\Big]
 \hspace{2mm}
 \Big[
 \ubarem \,  \gamma ^{\sigma} \, \big( \kslash + \qmoinsslash + m \big)
 \,  \gamma _{\mu} \, \vep 
 \Big] 
 \nonumber \\
\end{eqnarray}
\begin{eqnarray}\label{eq92}
D_2 = (k+\qmoins)^2 - m^2 = 2 \, k \cdot \qmoins
\end{eqnarray}
\begin{eqnarray}\label{eq93}
&&\hspace{-3mm}
 A^{\sigma} _3 ( \lambda_{e^+}, \lambda_{e^-} ; \lambda_{\bar{p}}, \lambda_p ) =
\frac{i}{q'^2} \ e^2_{p} \, e_{e^-} \
 \nonumber \\[1mm]
 &&
\Big[ 
\vbarpbar  \,
\Gamma_{NN \gamma }  ^{\sigma} (k) \
 \big( \kslash - \pmoinsslash + M \big)
 \,  \Gamma_{NN \gamma }  ^{\mu} (q') \,
\, \up  
\Big]
\nonumber \\[1mm]
&&
\hspace{0mm} 
\Big[    
\ubarem    \gamma _{\mu}  \vep
\Big]
\end{eqnarray}
\begin{eqnarray}\label{eq94}
D_3 = (\pmoins -k)^2 - M^2 = -2 \, k \cdot \pmoins
\end{eqnarray}
\begin{eqnarray}\label{eq95}
&&\hspace{-3mm}
 A^{\sigma} _4 ( \lambda_{e^+}, \lambda_{e^-} ; \lambda_{\bar{p}}, \lambda_p ) =
\frac{i}{q'^2} \ e^2_{p} \, e_{e^-} \
 \nonumber \\[1mm]
 &&
\Big[ 
\vbarpbar  \,
\Gamma_{NN \gamma }  ^{\mu} (q') \
 \big(\pplusslash - \kslash  + M \big) \,
\Gamma_{NN \gamma }  ^{\sigma} (k) \   \up  
\Big]
\nonumber \\[1mm]
&&
\hspace{0mm} 
\Big[    
\ubarem    \gamma _{\mu}  \vep
\Big]
\end{eqnarray}
\begin{eqnarray}\label{eq96}
D_4 = (\pplus -k)^2 - M^2= -2 \, k \cdot \pplus
\end{eqnarray}

\subsubsection{Cross section.}
 We define:
\begin{eqnarray}\label{eq97}
\overline{\Gamma} ^{\, \nu}_{NN \gamma } (x) = F^*_1 (x^2) \gamma ^{\nu} 
- \frac{F^*_2 (x^2)}{4 M } ( \gamma ^{\nu} \xslash - \xslash\gamma ^{\nu} )
\hspace{6mm}
\end{eqnarray}
\begin{eqnarray}\label{eq98}
\overline{\Gamma} ^{\, \sigma}_{NN \gamma } (k) = F_{10} (k^2) \gamma ^{\sigma} 
- \frac{F_{20} (k^2)}{4 M }
 ( \gamma ^{\sigma}  \kslash - \kslash \gamma ^{\sigma})
\hspace{6mm} 
\end{eqnarray}
\begin{eqnarray}\label{eq99}
\pmoinsslashM = \pmoinsslashmM 
\hspace{10mm} 
\pplusslashM = \pplusslashpM
\end{eqnarray}
\begin{eqnarray}\label{eq100}
\qmoinsslashm = \qmoinsslashpm
\hspace{10mm}
\qplusslashm = \qplusslashmm
\end{eqnarray}
Each $ X_{ij}$ term is the product of a constant times $ Y_{ij} $, the product  
of a hadronic tensor by a leptonic tensor. Each tensor is equal to
the trace of linear combination of the product of Dirac matrices:
\begin{eqnarray}\label{eq101}
X_{11} = -\frac{e_p ^2 \, e^4 _{e^-}}{q^4} \
&&  \hspace{-2mm}
Tr \Big\{
         \pmoinsslashM \,  \Gamma ^{\mu}_{NN \gamma } (q) \,
	 \pplusslashM  \, \overline{\Gamma} ^{\, \nu}_{NN \gamma } (q)
   \Big\}
   \nonumber  \\[1mm]
   && \hspace{-15mm}
Tr \Big\{
       \qmoinsslashm   \, \gamma _{\mu} (-\kslash - \qplusslashm )\,
       \gamma ^{\sigma} \, \qplusslashm  \, \gamma _{\sigma} \,
       (-\kslash - \qplusslashm ) \, \gamma _{\nu}
   \Big\}   
   \nonumber \\
\end{eqnarray}
\begin{eqnarray}\label{eq102}
X_{12} = -\frac{e_p ^2 \, e^4 _{e^-}}{q^4} \
&&  \hspace{-2mm}
Tr \Big\{
         \pmoinsslashM \,  \Gamma ^{\mu}_{NN \gamma } (q)\,
	 \pplusslashM \, \overline{\Gamma} ^{\, \nu}_{NN \gamma } (q)
   \Big\}
   \nonumber  \\[1mm]
   && \hspace{-22mm}
Tr \Big\{
       \qmoinsslashm \, \gamma _{\mu} (-\kslash - \qplusslashm )\,
       \gamma ^{\sigma} \,(\qplusslash - m  )\, \gamma _{\nu} \,
       (\kslash + \qmoinsslashm) \, \gamma _{\sigma}
   \Big\} 
      \nonumber \\  
\end{eqnarray}
\begin{eqnarray}\label{eq103}
X_{21} = -\frac{e_p ^2 \, e^4 _{e^-}}{q^4} \
&&  \hspace{-2mm}
Tr \Big\{
         \pmoinsslashM \,  \Gamma ^{\mu}_{NN \gamma } (q)\,
	 \pplusslashM \, \overline{\Gamma} ^{\, \nu}_{NN \gamma } (q)
   \Big\}
   \nonumber  \\[1mm]
   && \hspace{-22mm}
Tr \Big\{
       \qmoinsslashm \, \gamma ^{\sigma} \, (\kslash + \qmoinsslashm ) \, 
       \gamma _{\mu} \, \qplusslashm \, \gamma _{\sigma} \,
       (-\kslash - \qplusslashm ) \, \gamma _{\nu}
   \Big\}   
      \nonumber \\   
\end{eqnarray}
\begin{eqnarray}\label{eq104}
X_{22} = -\frac{e_p ^2 \, e^4 _{e^-}}{q^4} \
&&  \hspace{-2mm}
Tr \Big\{
         \pmoinsslashM \,  \Gamma ^{\mu}_{NN \gamma } (q)\,
	 \pplusslashM  \, \overline{\Gamma} ^{\, \nu}_{NN \gamma } (q)
   \Big\}
   \nonumber  \\[1mm]
   && \hspace{-22mm}
Tr \Big\{
       \qmoinsslashm \, \gamma ^{\sigma} \, (\kslash + \qmoinsslashm ) \, 
       \gamma _{\mu} \, \qplusslashm  \, \gamma _{\nu} \,
       (\kslash + \qmoinsslashm) \, \gamma _{\sigma}
   \Big\}   
      \nonumber \\   
\end{eqnarray}
\begin{eqnarray}\label{eq105}
X_{13} = -\frac{e_p ^3 \, e^3 _{e^-}}{q^2 \, q^{\prime ^{ \scriptstyle 2}}} \
&&  \hspace{-2mm}
Tr \Big\{
       \qmoinsslashm \, \gamma _{\mu} \, ( -\kslash - \qplusslashm )  \, 
       \gamma _{\sigma} \, \qplusslashm  \, \gamma _{\nu} \,
   \Big\}
   \nonumber  \\[1mm]
   && \hspace{-25mm}   
Tr \Big\{
         \pmoinsslashM \,  \Gamma ^{ \mu}_{NN \gamma } (q) \,
	 \pplusslashM \, \overline{\Gamma} ^{\, \nu}_{NN \gamma } (q') \,
	 ( \kslash - \pmoinsslashM)  \,  
	 \overline{\Gamma} ^{ \, \sigma} _{NN \gamma } (k)
   \Big\}
   \nonumber  \\      
\end{eqnarray}
\begin{eqnarray}\label{eq106}
X_{14} = -\frac{e_p ^3 \, e^3 _{e^-}}{q^2 \, q^{\prime ^{ \scriptstyle 2}}} \
&&  \hspace{-2mm}
Tr \Big\{
       \qmoinsslashm \, \gamma _{\mu} \, ( -\kslash - \qplusslashm ) \, 
       \gamma _{\sigma} \, \qplusslashm  \, \gamma _{\nu} \,
   \Big\}  
   \nonumber  \\[1mm]
   && \hspace{-25mm}
Tr \Big\{
         \pmoinsslashM \,  \Gamma ^{ \mu}_{NN \gamma } (q) \,
	 \pplusslashM \,
	 \overline{\Gamma} ^{\, \sigma } _{ NN \gamma } (k) \, 	 
	 (\pplusslashM - \kslash  ) \,
	  \overline{\Gamma} ^{\, \nu}_{NN \gamma } (q') 
   \Big\}
   \nonumber \\  
\end{eqnarray}
\begin{eqnarray}\label{eq107}
X_{23} = -\frac{e_p ^3 \, e^3 _{e^-}}{q^2 \, q^{\prime ^{ \scriptstyle 2}}} \
&&  \hspace{-2mm}
Tr \Big\{
       \qmoinsslashm  \, \gamma _{\sigma} \, ( \kslash + \qmoinsslashm ) \, 
       \gamma _{\mu} \, \qplusslashm  \, \gamma _{\nu} \,
   \Big\}
   \nonumber    \\[1mm]
   && \hspace{-25mm}
Tr \Big\{
         \pmoinsslashM \,  \Gamma ^{ \mu}_{NN \gamma } (q) \,
	 \pplusslashM  \, \overline{\Gamma} ^{\, \nu}_{NN \gamma } (q') \,
	 ( \kslash - \pmoinsslashM  ) \,  
	 \overline{\Gamma} ^{\, \sigma } _{NN \gamma } (k)
   \Big\}
   \nonumber    \\   
\end{eqnarray}
\begin{eqnarray}\label{eq108}
X_{24} = -\frac{e_p ^3 \, e^3 _{e^-}}{q^2 \, q^{\prime ^{ \scriptstyle 2}}} \
&&  \hspace{-2mm}
Tr \Big\{
       \qmoinsslashm  \, \gamma _{\sigma} \, ( \kslash + \qmoinsslashm ) \, 
       \gamma _{\mu}  \, \qplusslashm  \, \gamma _{\nu} \,
   \Big\} 
   \nonumber  \\[1mm]
   && \hspace{-25mm}
Tr \Big\{
         \pmoinsslashM \,  \Gamma ^{ \mu}_{NN \gamma } (q) \,
	 \pplusslashM  \, 
	 \overline{\Gamma} ^{\, \sigma } _{NN \gamma } (k) \,
	 (\pplusslashM - \kslash ) \,	 
	 \overline{\Gamma} ^{\, \nu}_{NN \gamma } (q') \,	 
   \Big\}
   \nonumber  \\  
\end{eqnarray}
\begin{eqnarray}\label{eq109}
X_{31} = -\frac{e_p ^3 \, e^3 _{e^-}}{q^2 \, q^{\prime ^{ \scriptstyle 2}}} \
&&  \hspace{-2mm}
Tr \Big\{
       \qmoinsslashm  \, \gamma _{\mu} \,  \qplusslashm   \, 
       \gamma _{\sigma} \, (-\kslash - \qplusslashm ) \, \gamma _{\nu} 
   \Big\}
   \nonumber  \\[1mm]
   && \hspace{-25mm}
Tr \Big\{
         \pmoinsslashM \, {\Gamma} ^{\, \sigma}_{NN \gamma } (k) \, 
	 ( \kslash - \pmoinsslashM ) \,
	 \Gamma ^{ \mu}_{NN \gamma } (q') \,
         \pplusslashM \,
         \overline{\Gamma} ^{\, \nu}_{NN \gamma } (q)	 
   \Big\}
   \nonumber  \\   
\end{eqnarray}
\begin{eqnarray}\label{eq110}
X_{32} = -\frac{e_p ^3 \, e^3 _{e^-}}{q^2 \, q^{\prime ^{ \scriptstyle 2}}} \
&&  \hspace{-2mm}
Tr \Big\{
       \qmoinsslashm  \, \gamma _{\mu} \,  \qplusslashm   \, 
       \gamma _{\nu} \, (\kslash + \qmoinsslashm  ) \, \gamma _{\sigma} 
   \Big\}
   \nonumber  \\[1mm]
   && \hspace{-25mm}
Tr \Big\{
         \pmoinsslashM \, {\Gamma} ^{\, \sigma}_{NN \gamma } (k) \, 
	 ( \kslash - \pmoinsslashM  ) \,
	 \Gamma ^{ \mu}_{NN \gamma } (q') \,
         \pplusslashM \,
         \overline{\Gamma} ^{\, \nu}_{NN \gamma } (q)	 
   \Big\}
   \nonumber  \\   
\end{eqnarray}
\begin{eqnarray}\label{eq111}
\hspace*{-12mm}
X_{33} = -\frac{e_p ^4 \, e^2 _{e^-}}{ q^{\prime ^{ \scriptstyle 4}}} \
&&  \hspace{-2mm}
Tr \Big\{
       \qmoinsslashm  \, \gamma _{\mu} \,  \qplusslashm   \, 
       \gamma _{\nu}  
   \Big\}
   \nonumber  \\[1mm]
   && \hspace{-25mm}
Tr \Big\{
         \pmoinsslashM \, {\Gamma} ^{\, \sigma}_{NN \gamma } (k) \, 
	 ( \kslash - \pmoinsslashM ) \,
	 \Gamma ^{ \mu}_{NN \gamma } (q') \,
         \pplusslashM \,
  \nonumber \\[1mm]
  &&	\hspace{-8mm} 
         \overline{\Gamma} ^{\, \nu}_{NN \gamma } (q') \, 	 
         ( \kslash - \pmoinsslashM  ) \
         \overline{\Gamma} _{ \, \sigma NN \gamma } (k) \
   \Big\}   
\end{eqnarray}
\begin{eqnarray}\label{eq112}
\hspace*{-12mm}
X_{34} = -\frac{e_p ^4 \, e^2 _{e^-}}{ q^{\prime ^{ \scriptstyle 4}}} \
&&  \hspace{-2mm}
Tr \Big\{
       \qmoinsslashm  \, \gamma _{\mu} \,  \qplusslashm  \, 
       \gamma _{\nu}  
   \Big\}
   \nonumber  \\[1mm]
   && \hspace{-25mm}
Tr \Big\{
         \pmoinsslashM \, {\Gamma} ^{\, \sigma}_{NN \gamma } (k) \, 
	 ( \kslash - \pmoinsslashM ) \,
	 \Gamma ^{ \mu}_{NN \gamma } (q') \,
         \pplusslashM \,
  \nonumber \\[1mm]
  &&	\hspace{-8mm} 	 	 
         \overline{\Gamma} _{ \, \sigma NN \gamma } (k)	 \,
	 (\pplusslashM - \kslash ) \	 
         \overline{\Gamma} ^{\, \nu}_{NN \gamma } (q') \ 	 
   \Big\}   
\end{eqnarray}
\begin{eqnarray}\label{eq113}
X_{41} = -\frac{e_p ^3 \, e^3 _{e^-}}{q^2 \, q^{\prime ^{ \scriptstyle 2}}} \
&&  \hspace{-2mm}
Tr \Big\{
       \qmoinsslashm  \, 
       \gamma _{\mu} \,  
       \qplusslashm  \, 
       \gamma _{\sigma} \, 
       ( -\kslash - \qplusslashm )
       \gamma _{\nu} 
   \Big\}
   \nonumber  \\[1mm]
   && \hspace{-25mm}
Tr \Big\{
         \pmoinsslashM \, 
	 \Gamma ^{ \mu}_{NN \gamma } (q') \,
	 (\pplusslashM - \kslash  ) \,
	 {\Gamma} ^{\, \sigma}_{NN \gamma } (k) \, 
	 \pplusslashM  \,         
         \overline{\Gamma} ^{\, \nu}_{NN \gamma } (q)	 
   \Big\}
   \nonumber  \\  
\end{eqnarray}
\begin{eqnarray}\label{eq114}
X_{42} = -\frac{e_p ^3 \, e^3 _{e^-}}{q^2 \, q^{\prime ^{ \scriptstyle 2}}} \
&&  \hspace{-2mm}
Tr \Big\{
       \qmoinsslashm  \, 
       \gamma _{\mu} \, 
        \qplusslashm   \, 
       \gamma _{\nu} \, 
       (\kslash + \qmoinsslashm) \,
        \gamma _{\sigma} 
   \Big\}
   \nonumber  \\[1mm]
   && \hspace{-25mm}
Tr \Big\{
         \pmoinsslashM \, 
          \Gamma ^{ \mu}_{NN \gamma } (q') \,
         (\pplusslashM - \kslash  ) \,
         {\Gamma} ^{\, \sigma}_{NN \gamma } (k) \,
         \pplusslashM \,
	 \overline{\Gamma} ^{\, \nu}_{NN \gamma } (q)	 
   \Big\}
   \nonumber  \\  
\end{eqnarray}
\begin{eqnarray}\label{eq115}
\hspace*{-12mm}
X_{43} = -\frac{e_p ^4 \, e^2 _{e^-}}{ q^{\prime ^{ \scriptstyle 4}}} \
&&  \hspace{-2mm}
Tr \Big\{
       \qmoinsslashm  \, \gamma _{\mu} \,  \qplusslashm   \, 
       \gamma _{\nu}  
   \Big\}   
   \nonumber  \\[1mm]
   && \hspace{-25mm}
Tr \Big\{
         \pmoinsslashM \, 
          \Gamma ^{ \mu}_{NN \gamma } (q') \,
          (\pplusslashM - \kslash ) \,
	 {\Gamma} ^{\, \sigma}_{NN \gamma } (k) \, 
         \pplusslashM \,
  \nonumber \\[1mm]
  &&	\hspace{-8mm}	 
         \overline{\Gamma} ^{\, \nu}_{NN \gamma } (q') \,
         ( \kslash - \pmoinsslashM ) \,
         \overline{\Gamma} _{ \, \sigma NN \gamma } (k)	 \,	
   \Big\}
\end{eqnarray}
\begin{eqnarray}\label{eq116}
\hspace*{-12mm}
X_{44} = -\frac{e_p ^4 \, e^2 _{e^-}}{ q^{\prime ^{ \scriptstyle 4}}} \
&&  \hspace{-2mm}
Tr \Big\{
       \qmoinsslashm  \, \gamma _{\mu} \,  \qplusslashm   \, 
       \gamma _{\nu}  
   \Big\}
   \nonumber  \\[1mm]
   && \hspace{-22mm}
Tr \Big\{
         \pmoinsslashM \, 
          \Gamma ^{ \mu}_{NN \gamma } (q') \,
          (\pplusslashM - \kslash ) \,
	 {\Gamma} ^{\, \sigma}_{NN \gamma } (k) \, 
         \pplusslashM \,
  \nonumber \\[1mm]
  &&	\hspace{-8mm}	 
         \overline{\Gamma} _{ \, \sigma NN \gamma } (k)	 \,
         (\pplusslashM - \kslash ) \,
         \overline{\Gamma} ^{\, \nu}_{NN \gamma } (q') 	
   \Big\}   
\end{eqnarray}
When the photon is emitted by the hadrons or by the leptons, the hadronic
and the leptonic tensors contain ten terms. The product of the traces is
written
\begin{eqnarray}\label{eq117}
Y_{ij} = \sum_{\mathrm{k}, \mathrm{k}' =1}^{10}
                C^h_{ij \mathrm{k} } \, 
                C^{\ell}_{ij \mathrm {k}'} \, 
T^{\mu \nu  }_{_{\scriptstyle H}}(\mathrm{k}) \, 
T_{_{\scriptstyle L} \, _{_{\scriptstyle  \mu \nu }}} (\mathrm{k}')
\end{eqnarray}
The tensors $ T^{\mu \nu  }_{_{\scriptstyle H}}(\mathrm{k}) $,
$ T_{_{\scriptstyle L} \, _{_{\scriptstyle  \mu \nu }}}(\mathrm{k}')$
are given in Table \ref{tablen}
\vspace{-2mm}
\begin{table}[H]
\caption{Hadronic and Leptonic tensors I}
\label{tablen}
\begin{center}
\begin{tabular}{|rc|rc|}
\hline\noalign{\smallskip}
\ k  & \  $ T^{\mu \nu  }_{_{\scriptstyle H}}   $ & \ \ k' 
   &    $  T_{_{\scriptstyle L} \, _{_{\scriptstyle  \mu \nu }}}  $                    \\
\noalign{\smallskip}\hline\noalign{\smallskip}
\  1   & \  $  g^{\mu \nu}              $ & 1 & \ $  g_{\mu \nu}                 $ \\

\  2   & \  $  k^{\mu} \, k^{\nu}       $ & 2 & \ $  k_{\mu} \, k_{\nu}          $ \\
\  3   & \  $  k^{\mu} \, p^{-\nu}      $ & 3 & \ $  k_{\mu} \, q^{+}_{\nu}      $ \\
\  4   & \  $  k^{\mu} \, p^{+\nu}      $ & 4 & \ $  k_{\mu} \, q^{-}_{\nu}      $ \\

\  5   & \  $  p^{-\mu} \, k^{\nu}      $ & 5 & \ $  q^{+}_{\mu} \, k_{\nu}      $ \\
\  6   & \  $  p^{-\mu} \, p^{-\nu}     $ & 6 & \ $  q^{+}_{\mu} \, q^{+}_{\nu}  $ \\
\  7   & \  $  p^{-\mu} \, p^{+\nu}     $ & 7 & \ $  q^{+}_{\mu} \, q^{-}_{\nu}  $ \\

\  8   & \  $  p^{+\mu} \, k^{\nu}      $ & 8 & \ $  q^{-}_{\mu} \, k_{\nu}      $ \\
\  9   & \  $  p^{+\mu} \, p^{-\nu}     $ & 9 & \ $  q^{-}_{\mu} \, q^{+}_{\nu}  $ \\
\ 10   & \  $  p^{+\mu} \, p^{+\nu}     $ & 10 & \ $  q^{-}_{\mu} \, q^{-}_{\nu}  $ \\
\noalign{\smallskip}\hline
\end{tabular}
\end{center}
\vspace*{-5mm}
\end{table}
When the photon is emitted by a hadron and by a lepton, the hadronic and 
the leptonic tensors contain thirty six terms. 
\begin{eqnarray}\label{eq118}
Y_{ij} = \sum_{\mathrm{k}, \mathrm{k}' =1}^{36}
                C^h_{ij \mathrm{k} } \, 
                C^{\ell}_{ij \mathrm {k}'} \, 
T^{\mu \nu  }_{_{\scriptstyle H}}(\mathrm{k}) \, 
T_{_{\scriptstyle L} \, _{_{\scriptstyle  \mu \nu }}} (\mathrm{k}')
\end{eqnarray}
with the tensors 
$ T^{\mu \nu  }_{_{\scriptstyle H}}(\mathrm{k}) $,
$ T_{_{\scriptstyle L} \, _{_{\scriptstyle  \mu \nu }}}(\mathrm{k}')$
given in Table \ref{tablenpun}.
Each coefficient $ C^h_{ij \mathrm{k} } $
in the hadronic tensor depends on the electromagnetic form factors
and its analytical expression has been derived with the help of Mathematica [28],
but will not be given in this paper. A numerical check, obtained in calculating
 numerically the hadronic and leptonic traces, has been done.

\vspace{-3mm}
\begin{table}[H]
\caption{Hadronic and Leptonic tensors II}
\label{tablenpun}
\begin{center}
\begin{tabular}{|rc|rc|}
\hline\noalign{\smallskip}
\ k  & \  $ T^{\mu \nu \sigma }_{_{\scriptstyle H}}  $ & \ \ k'  
&   $  T_{_{\scriptstyle L} \, _{_{\scriptstyle  \mu \nu \sigma}}} $                    \\
\noalign{\smallskip}\hline\noalign{\smallskip}
\  1   & \  $  g^{\mu \nu} \, k^{\sigma}             $ & 1 & \ $  g_{\mu \nu} \, k_{\sigma}                $ \\
\  2   & \  $  g^{\mu \nu} \, p^{-\sigma}            $ & 2 & \ $  g_{\mu \nu} \, q^{+}_{\sigma}            $ \\
\  3   & \  $  g^{\mu \nu} \, p^{+\sigma}            $ & 3 & \ $  g_{\mu \nu} \, q^{-}_{\sigma}            $ \\
\  4   & \  $  g^{\mu \sigma} \, k^{\nu}             $ & 4 & \ $  g_{\mu \sigma} \, k_{\nu}                $ \\
\  5   & \  $  g^{\mu \sigma} \, p^{-\nu}            $ & 5 & \ $  g_{\mu \sigma} \, q^{+}_{\nu}            $ \\
\  6   & \  $  g^{\mu \sigma} \, p^{+\nu}            $ & 6 & \ $  g_{\mu \sigma} \, q^{-}_{\nu}            $ \\
\  7   & \  $  g^{\nu \sigma} \, k^{\mu}             $ & 7 & \ $  g_{\nu \sigma} \, k_{\mu}                $ \\
\  8   & \  $  g^{\nu \sigma} \, p^{-\mu}            $ & 8 & \ $  g_{\nu \sigma} \, q^{+}_{\mu}            $ \\
\  9   & \  $  g^{\nu \sigma} \, p^{+\mu}            $ & 9 & \ $  g_{\nu \sigma} \, q^{-}_{\mu}            $ \\
\ 10   & \  $  k^{\mu} \, k^{\nu} \, k^{\sigma}      $ & 10 & \ $  k_{\mu} \, k_{\nu} \, k_{\sigma}         $ \\
\ 11   & \  $  k^{\mu} \, k^{\nu} \, p^{-\sigma}     $ & 11 & \ $  k_{\mu} \, k_{\nu} \, q^{+}_{\sigma}     $ \\
\ 12   & \  $  k^{\mu} \, k^{\nu} \, p^{+\sigma}     $ & 12 & \ $  k_{\mu} \, k_{\nu} \, q^{-}_{\sigma}     $ \\
\ 13   & \  $  k^{\mu} \, p^{-\nu} \, k^{\sigma}     $ & 13 & \ $  k_{\mu} \, q^{+}_{\nu} \, k_{\sigma}     $ \\
\ 14   & \  $  k^{\mu} \, p^{-\nu} \, p^{-\sigma}    $ & 14 & \ $  k_{\mu} \, q^{+}_{\nu} \, q^{+}_{\sigma} $ \\
\ 15   & \  $  k^{\mu} \, p^{-\nu} \, p^{+\sigma}    $ & 15 & \ $  k_{\mu} \, q^{+}_{\nu} \, q^{-}_{\sigma} $ \\
\ 16   & \  $  k^{\mu} \, p^{+\nu} \, k^{\sigma}     $ & 16 & \ $  k_{\mu} \, q^{-}_{\nu} \, k_{\sigma}     $ \\
\ 17   & \  $  k^{\mu} \, p^{+\nu} \, p^{-\sigma}    $ & 17 & \ $  k_{\mu} \, q^{-}_{\nu} \, q^{+}_{\sigma} $ \\
\ 18   & \  $  k^{\mu} \, p^{+\nu} \, p^{+\sigma}    $ & 18 & \ $  k_{\mu} \, q^{-}_{\nu} \, q^{-}_{\sigma} $ \\
\ 19   & \  $  p^{-\mu} \, k^{\nu} \, k^{\sigma}     $ & 19 & \ $  q^{+}_{\mu} \, k_{\nu} \, k_{\sigma}         $ \\
\ 20   & \  $  p^{-\mu} \, k^{\nu} \, p^{-\sigma}    $ & 20 & \ $  q^{+}_{\mu} \, k_{\nu} \, q^{+}_{\sigma}     $ \\
\ 21   & \  $  p^{-\mu} \, k^{\nu} \, p^{+\sigma}    $ & 21 & \ $  q^{+}_{\mu} \, k_{\nu} \, q^{-}_{\sigma}     $ \\
\ 22   & \  $  p^{-\mu} \, p^{-\nu} \, k^{\sigma}    $ & 22 & \ $  q^{+}_{\mu} \, q^{+}_{\nu} \, k_{\sigma}     $ \\
\ 23   & \  $  p^{-\mu} \, p^{-\nu} \, p^{-\sigma}   $ & 23 & \ $  q^{+}_{\mu} \, q^{+}_{\nu} \, q^{+}_{\sigma} $ \\
\ 24   & \  $  p^{-\mu} \, p^{-\nu} \, p^{+\sigma}   $ & 24 & \ $  q^{+}_{\mu} \, q^{+}_{\nu} \, q^{-}_{\sigma} $ \\
\ 25   & \  $  p^{-\mu} \, p^{+\nu} \, k^{\sigma}    $ & 25 & \ $  q^{+}_{\mu} \, q^{-}_{\nu} \, k_{\sigma}     $ \\
\ 26   & \  $  p^{-\mu} \, p^{+\nu} \, p^{-\sigma}   $ & 26 & \ $  q^{+}_{\mu} \, q^{-}_{\nu} \, q^{+}_{\sigma} $ \\
\ 27   & \  $  p^{-\mu} \, p^{+\nu} \, p^{+\sigma}   $ & 27 & \ $  q^{+}_{\mu} \, q^{-}_{\nu} \, q^{-}_{\sigma} $ \\
\ 28   & \  $  p^{+\mu} \, k^{\nu} \, k^{\sigma}     $ & 28 & \ $  q^{-}_{\mu} \, k_{\nu} \, k_{\sigma}         $ \\
\ 29   & \  $  p^{+\mu} \, k^{\nu} \, p^{-\sigma}    $ & 29 & \ $  q^{-}_{\mu} \, k_{\nu} \, q^{+}_{\sigma}     $ \\
\ 30   & \  $  p^{+\mu} \, k^{\nu} \, p^{+\sigma}    $ & 30 & \ $  q^{-}_{\mu} \, k_{\nu} \, q^{-}_{\sigma}     $ \\
\ 31   & \  $  p^{+\mu} \, p^{-\nu} \, k^{\sigma}    $ & 31 & \ $  q^{-}_{\mu} \, q^{+}_{\nu} \, k_{\sigma}     $ \\
\ 32   & \  $  p^{+\mu} \, p^{-\nu} \, p^{-\sigma}   $ & 32 & \ $  q^{-}_{\mu} \, q^{+}_{\nu} \, q^{+}_{\sigma} $ \\
\ 33   & \  $  p^{+\mu} \, p^{-\nu} \, p^{+\sigma}   $ & 33 & \ $  q^{-}_{\mu} \, q^{+}_{\nu} \, q^{-}_{\sigma} $ \\
\ 34   & \  $  p^{+\mu} \, p^{+\nu} \, k^{\sigma}    $ & 34 & \ $  q^{-}_{\mu} \, q^{-}_{\nu} \, k_{\sigma}     $ \\
\ 35   & \  $  p^{+\mu} \, p^{+\nu} \, p^{-\sigma}   $ & 35 & \ $  q^{-}_{\mu} \, q^{-}_{\nu} \, q^{+}_{\sigma} $ \\
\ 36   & \  $  p^{+\mu} \, p^{+\nu} \, p^{+\sigma}   $ & 36 & \ $  q^{-}_{\mu} \, q^{-}_{\nu} \, q^{-}_{\sigma} $ \\
\noalign{\smallskip}\hline
\end{tabular}
\end{center}
\vspace*{0cm}
\end{table}

\subsection{Soft photon and factorization.}
In this section, we show how the soft  photon cross section factorizes
in term of the Born cross section.

The product of the hadronic and leptonic tensors in the Born term
is given by
\begin{eqnarray}\label{eq119}
\hspace*{-5mm}
Y_{00} = 
&&  \hspace{-1mm}
Tr \Big\{
         \pmoinsslashM \,  \Gamma ^{\mu}_{NN \gamma } (q) \,
	 \pplusslashM  \, \overline{\Gamma} ^{\, \nu}_{NN \gamma } (q)
   \Big\}
   \nonumber  \\[1mm]
   && \hspace{-1mm}
Tr \Big\{
       \qmoinsslashm  \, \gamma _{\mu} \,  \qplusslashm   \, 
       \gamma _{\nu}  
   \Big\}   
\end{eqnarray}
When the  photon energy goes to zero, we have: 
\begin{eqnarray}
&&
\Gamma ^{\sigma}_{NN \gamma } (k) 
 \rightarrow  \gamma ^{\sigma}; 
 \hspace{5mm}
\overline{\Gamma} ^{\, \sigma}_{NN \gamma } (k) \rightarrow \gamma ^{\sigma}
\label{eq120} \\[1mm]
&&
 q' = \pmoins + \pplus - k = q -k \rightarrow q
\label{eq121} \\[1mm] 
&&
Y_{ij} \rightarrow Y_{ij} ^{\soft}
\label{eq122}
\end{eqnarray}
We have seen in eq.(\ref{eq44}) that there are ten terms
which have a well definite limit when the photon energy goes to zero.
So the factorization of the soft cross section in terms of the 
Born cross section can be studied
through the quantities $ - {Y ^{\soft}_{11}}/{4 \, Y_{00}} $ to
$ -(Y ^{\soft}_{34} + Y ^{\soft}_{43})/{4 \, Y_{00}}  $. In these
quantities, the numerator and the denominator depend on the 
electromagnetic form factors. The derivation of these
quantities has been done with the help of Mathematica [28].
The result is given in table \ref{tablenpdeux}.
\begin{table}[H]
\begin{center}
\caption{Ratio of soft photon cross section to Born cross section}
\label{tablenpdeux}
\begin{tabular}{|c|r|}
\hline
\ Definition                                 &        Result                 \\
\hline
$-{Y ^{\soft}_{11}}/{4 \, Y_{00}}           $&$ -{\qplus } ^2                $\\
$-{Y ^{\soft}_{22}}/{4 \, Y_{00}}           $&$ -{\qmoins} ^2                $\\
$-{Y ^{\soft}_{33}}/{4 \, Y_{00}}           $&$ -{\pmoins }^2                $\\ 
$-{Y ^{\soft}_{44}}/{4 \, Y_{00}}           $&$ -{\pplus  }^2                $\\
$-({Y ^{\soft}_{12} + Y ^{\soft}_{21}})/{4 \, Y_{00}} \hspace{7mm }   $&$  2 \qplus  \cdot \qmoins  $\\
$-({Y ^{\soft}_{13} + Y ^{\soft}_{31}})/{4 \, Y_{00}} \hspace{7mm }   $&$ -2 \qplus  \cdot \pmoins  $\\ 
$-({Y ^{\soft}_{14} + Y ^{\soft}_{41}})/{4 \, Y_{00}} \hspace{7mm }   $&$  2 \qplus  \cdot  \pplus  $\\
$-({Y ^{\soft}_{23} + Y ^{\soft}_{32}})/{4 \, Y_{00}} \hspace{7mm }   $&$  2 \qmoins \cdot \pmoins  $\\
$-({Y ^{\soft}_{24} + Y ^{\soft}_{42}})/{4 \, Y_{00}} \hspace{7mm }   $&$ -2 \qmoins \cdot \pplus   $\\
$-({Y ^{\soft}_{34} + Y ^{\soft}_{43}})/{4 \, Y_{00}} \hspace{7mm }   $&$  2 \pmoins \cdot \pplus   $\\
\hline
\end{tabular}
\end{center}
\end{table}
The result is very interesting. It shows that each ratio is 
independent of the electromagnetic form factors.
The sum of each term of the right column of the Table 6
divided by its corresponding propagator lead to the 
 following analytical function:
\begin{eqnarray}\label{eq123}
\hspace*{-5mm}
F^{\soft} = 
-\bigg(
     \frac{p^{+}}{k.p^{+}} - \frac{p^{-}}{k.p^{-}}
   + \frac{q^{-}}{k.q^{-}} - \frac{q^{+}}{k.q^{+}}
  \, \bigg)^2  
\end{eqnarray}
The soft correction $ \delta _\soft$ is obtained through 
the integration of this function over the photon variables
(Eq.23).

\end{appendix}

\end{document}